\documentclass[lettersize,journal]{IEEEtran}
\usepackage{amsmath,amsfonts}
\usepackage{algorithmicx,algorithm}
\usepackage{array}
\usepackage{textcomp}
\usepackage{stfloats}
\usepackage{url}
\usepackage{verbatim}
\usepackage{graphicx}
\usepackage{multirow}
\usepackage{subfigure}
\usepackage{ulem}
\usepackage{diagbox}
\usepackage{multirow}
\usepackage{multicol,multicap,graphicx}
\usepackage{caption}
\usepackage{lineno} 
\usepackage[explicit]{titlesec}
\usepackage{setspace}
\usepackage{color}
\usepackage{cite}
\usepackage{amssymb}
\usepackage{amsthm,amsmath,amssymb}
\usepackage{bbm}
\usepackage{bm}
\usepackage{mathrsfs}
\usepackage{float}
\usepackage{CJK}

\def\BibTeX{{\rm B\kern-.05em{\sc i\kern-.025em b}\kern-.08em
    T\kern-.1667em\lower.7ex\hbox{E}\kern-.125emX}}
\usepackage{balance}
\begin{document}

\title{
  Resource Allocation in Cell-Free MU-MIMO Multicarrier System with Finite and Infinite Blocklength}
  \author{Jiafei Fu, Pengcheng Zhu,~\IEEEmembership{Member,~IEEE}, Bo Ai,~\IEEEmembership{Fellow,~IEEE}, \\Jiangzhou Wang,~\IEEEmembership{Fellow,~IEEE}, Xiaohu You,~\IEEEmembership{Fellow,~IEEE}
\thanks{Jiafei Fu, Pengcheng Zhu and Xiaohu You are with National Mobile Communications Research Laboratory, Southeast University, Nanjing 210096,
China. (e-mails: 
fujfei@seu.edu.cn; p.zhu@seu.edu.cn; xhyu@seu.edu.cn). 
Bo Ai is with the State Key Laboratory of Rail Traffic
Control and Safety, Beijing Jiaotong University, Beijing 100044, China. (e-mail: boai@bjtu.edu.cn).
Jiangzhou Wang is with the School of Engineering, University of Kent, CT2 7NZ Canterbury, U.K. (e-mail: j.z.wang@kent.ac.uk). Corresponding author: Pengcheng Zhu.}}

\markboth{Journal of \LaTeX\ Class Files,~Vol.~18, No.~9, September~2020}%
{How to Use the IEEEtran \LaTeX \ Templates}

\maketitle

\begin{abstract}
  The explosive growth of data in the next generation mobile communications results in more scarce spectrum resources. It is important to optimize the system performance under limited resources.
  In this paper, we investigate how to achieve weighted throughput (WTP) maximization for cell-free (CF) multiuser MIMO (MU-MIMO) multicarrier (MC) systems through resource allocation (RA), in the cases of finite blocklength (FBL) and infinite blocklength (INFBL) regimes. 
  To ensure the quality of service (QoS) of each user, particularly for the block error rate (BLER) and latency in the FBL regime, the WTP gets maximized under the constraints of total power consumption and required QoS metrics. 
  Since the channels vary in different subcarriers (SCs) and inter-user interference strengths, the WTP can be maximized by scheduling the best users in each time-frequency (TF) resource and advanced beamforming design, while the resources can be fully utilized. 
  With this motivation, we introduce a 0-1 scheduling indication variable to represent whether the user is selected on the current TF resource. 
  However, the problem mentioned above is a mixed integer nonlinear programming (MINLP) problem that is challenging to address due to the integer scheduling indication variable and the beamforming vector. Therefore, we propose a nested iteration algorithm to solve this highly non-convex problem, where the user scheduling (US) scheme is optimized by gene-aided (GA) algorithm in the outer iteration, while the successive convex approximation (SCA) is used for beamforming design in inner iteration. 
  Simulation results demonstrate that the proposed MU-MIMO scheduling scheme outperforms the other two traditional scheduling schemes, including single-user MIMO (SU-MIMO) and full-user MIMO (FU-MIMO) RA schemes that may not necessarily be optimal. 
  And the CF system in our scenario is capable of achieving higher spectral efficiency (SE) than the centralized antenna systems (CAS).
\end{abstract}


\section{Introduction}
\IEEEPARstart{T}{o} meet the requirements of the explosive growth of the future mobile communication services, researchers focused on the research and development of the next generation mobile communications to achieve global coverage and interconnection of all things.
Technologies with higher rates, wider coverage, higher connection density, ultra-reliability, and low-latency (URLLC) are studied and developed \cite{ref_you_6gECES}, so as to satisfy the various needs of communication services for vertical applications, such as autonomous driving, factory automation, and remote surgery \cite{ref_CUIFE,ref_MLRRS}. 

Cell-free (CF) as a novel cellular architecture without cell boundary can provide consistent services for all users with coordinated multipoint (CoMP) transmission \cite{cellfreeElhoushy,cellfreeMM}. 
In cell-free systems, multiple access points (APs) connected to a central processor unit (CPU) in a baseband pool (BBP) through a wireless backhaul link or optical fiber collaboratively transmit signals to all users. The channel state information (CSI) is shared among the APs. Data-sharing and compression strategies are two commonly used transmission strategies to account for the limited backhaul capacity.
For the data-sharing, APs cooperatively transmit the beamformed signals formed locally to the users.
For the compression, the beamforming operation is performed in BBP first. The CPU compresses and forwards the beamformed signals to the APs, where compression is needed due to finite backhaul capacity. 
The extra DoF offered by distributed antenna cooperation can improve the spectral efficiency significantly \cite{ref_JUSTX}.
Besides, the distance between APs and equipment becomes closer, reducing the transmission latency. 
Hence, CF is a potential paradigm that can be applied in short-packet transmission to enhance ultra-reliability and low-latency communications  (URLLC).

The short-packet structure is considered as the typical frame structure in URLLC to realize the stringent requirements of 100us latency and $10^{-6}$ block error rate (BLER) \cite{REF_you_T6CNV}, which is different from the Shannon theory where BLER and the packet length are assumed to be infinitely small and large \cite{ref_spt_JF}, respectively.
In \cite{ref_CCRF}, the tight approximation of the maximal achievable rate was first obtained by Polyanskiy at a given blocklength and the block error rate, which characterizes the relationship among rate, blocklength, and block error rate. This groundbreaking research has laid the foundation for the short-packet communication theory, thereby providing technical support for vertical industry applications. Yang in \cite{spcYang2014} further established achievability and converse bounds on the maximal achievable rate in quasi-static MIMO fading channels. According to the characteristics of short-packet communications \cite{spcYang2014,spc1,spc2}, the outage probability concept considering both security and reliability with an eavesdropper was proposed in \cite{ref_RSSPC}, while effective outage capacity was further established as the performance metric to evaluate the security and reliability. The author in \cite{ref_ASBA} studied the average secure block error rate of the downlink NOMA  system in consideration of short-packet communications in flat Rayleigh fading channels. Besides, the author in \cite{ref_EESP} investigated the energy efficiency (EE) for non-orthogonal multiple access (NOMA) massive machine-type communication (mMTC) networks by optimizing the subchannel and power control, where sporadic and low-rate short-packet were used for information exchange among mMTC devices.
As the number of users increases, how to satisfy the QoS of each user in mMTC with limited resources is still an open problem.

Furtunately, multicarrier transmission (MCT) (e.g., orthogonal frequency division multiple access (OFDMA)) is applied to guarantee high data rate transmission due to its flexibility for resources allocation, its resistance to multi-path fading and its ability to realize the multi-user diversity \cite{jw1,jw2,ref_EERAOS}. 
The author in \cite{Sun_2018} studied the resource allocation in multiple-input single-output (MISO) multicarrier (MC) NOMA systems, in which a full-duplex (FD) base station serves multiple half-duplex (HD) uplink and downlink users on the same subcarrier simultaneously, to maximize the WTP. An optimal power and subcarrier allocation scheme was proposed in \cite{Yan Sun 2017} to provide a good balance between improving the system throughput and maintaining fairness among users. Compared to the traditional single-user MIMO (SU-MIMO) \cite{802.11n} in MCT systems 
that only one best user is likely to be served at each time-frequency resource and may not be able to fully utilize the system resources, MU-MIMO in MCT systems can provide additional degrees of freedom (DoF) to obtain diversity and multiplexing gains, where multiple single-antenna users are served by a transmitter equipped with a large number of antennas \cite{ref_multiuserMIMO,mu-mimo1,mu-mimo2,mu-mimo3}.

To this end, the author in \cite{He2021} investigated three optimization problems regarding the weighted sum rate maximization, EE maximization, and user fairness problems by beamforming design for the downlink multiuser URLLC system, in which spatial multiplexing gain at the base station can be achieved without the need for multiple-antenna users. In \cite{LiuBo2023}, the author proposed a low-complexity algorithm to optimize the beamformer and the remote radio unit selection for improving the EE of a distributed massive MIMO multiuser system.
However, since all users are scheduled in each time-frequency resource (Here, we define this type of multiuser scheduling MIMO as full-user MIMO (FU-MIMO)), these MC systems do not have user scheduling (US) operations actrually.  
Since the channel fading coefficients at different subcarriers are statistically independent for different users, the maximal capacity can be obtained by choosing the best users for each subcarrier and allocating the corresponding transmitting power. 

In order to investigate the user scheduling gain in MU-MIMO system, some studies have been carried out related to both resource allocation and multi-user scheduling in each subcarrier.
The author in \cite{ref_ORALMO} explored joint uplink-downlink resource allocation for multi-user OFDMA-URLLC in mobile edge computing (MEC) systems. The problem regarding the minimization of the total weighted power consumption of the system was studied under the constraint of the end-to-end delay of a computation task of each user.
\cite{ref_JBPSO} formulated an optimization problem for maximizing the weighted system sum throughput while guaranteeing the QoS of the URLLC users in intelligent reflecting surface (IRS) enhanced URLLC multicell networks with OFDM access. 
Cheng researched a robust resource block assignment and beamforming design problem for minimizing the total transmit power of the central controller in OFDMA-URLLC system about the worst-case SNR while guaranteeing URLLC requirements \cite{ref_RRAMAU}. 
As shown in \cite{Coordinated2021}, the network utility maximization problem is considered by optimizing the power vector and user selection for SU-MIMO and MU-MIMO in a coordinated massive MIMO system. 
The author in \cite{Dynamic2022} constructed a joint system throughput and user rate satisfation problem in CF massive MIMO systems. A dynamic user scheduling algorithm was proposed to adjust the list and number of paired users according to the channel quality and transmission requirements of users.

However, the aforementioned studies usually concentrate on  joint user scheduling and resource allocation in multi-cell or single base station scenarios \cite{multicell1,multicell2,multicell3}. Most papers discussed resource allocation while full-user scheduling is considered to obtain diversity gain in CF massive MIMO system. Few works paid attention to CF MU-MIMO resource allocation. 
Therefore, we allocate the resources in CF MU-MIMO MC systems with both infinite blocklength (INFBL) and finite blocklength (FBL). The centralized antenna system (CAS) is also investigated as the comparison scheme. 
Different from the single-user scheduling and full-user scheduling scenarios, we investigate the weighted throughput (WTP) in terms of the multi-user scheduling and the beamforming design in CF systems. In the sequal, we propose a nested iterative algorithm named USBDA to solve an MINLP problem. Furthermore, a two-stage algorithm is used for the USBDA to reduce the computational complexity. 
The main contributions of this work are listed below. 
\begin{itemize}
\item We consider a WTP maximization problem in a CF MU-MIMO MC system by optimizing the user scheduling scheme and the beamforming vector while satisfying the constraints of the total power consumption and the minimum QoS of each user with given TF resources and BLER. The CAS is investigated as the comparison scheme of CF systems in our proposed scenarios. And the WTP maximization problem is applied in both INFBL and FBL regimes.   
\item By beamforming design and scheduling the multiple best users according to the channel condition in different frequency resources, the WTP can be maximized and the limited resources can be fully utilized in MU-MIMO scenario compared to the traditional resource allocation in SU-MIMO and FU-MIMO scenarios.
The user scheduling optimization problem is transformed to optimize a 0-1 scheduling indication variable that represents whether the user is scheduled on the current time-frequency resource. Hence each user's blocklength equals the sum of its corresponding scheduling indication variables in the total radio resources. 
\item A nested iterative algorithm named USBDA is proposed to solve the WTP maximization problem both in FBL and INFBL regimes respectively. The inner iteration is the beamforming design, and the gene-aided (GA) algorithm in the outer iteration is proposed for user scheduling optimization.  
Two-stage algorithm is proposed for the USBDA algorithm to reduce the computational complexity. For the first stage, the linear minimum mean square error (MMSE) beamformer rather than the proposed beamformer is used in the USBDA algorithm to obtain a preferable initial value of the US scheme. Hence, fewer generations are needed in the USBDA algorithm with the proposed beamformer in the second stage. Simulation results verify this advantage.

\end{itemize}

The rest of this paper is organized as follows. Section II describes the system model and formulates the WTP expression in CF MU-MIMO MC system both in INFBL regime and FBL regime with given block error rate. Section III and IV illustrate the solutions in INFBL and FBL regimes to solve the maximization problem of the WTP by optimizing the user scheduling scheme and the beamformer, respectively. And the proposed problem is applied to CAS in Section IV. Numerical results are given in Section V to compare the difference between CAS and CF systems. And demonstrate the effectiveness of the proposed MU-MIMO resource allocation scheme compared to its counterparts. Section VI concludes this paper.

Notation: Boldface lower-case letter, boldface upper-case letter, and lower-case letter represent a vector, matrix, and scaler respectively. 
$(\cdot)^T$, $(\cdot)^H$ denotes the transpose and the conjugate transpose.
The expectation and the trace operation are denoted as $\mathbb{E}\{\cdot\}$ and 
${\textrm{Tr}}\{\cdot\}$ respectively.
Calligraphy letters $\mathcal{M}$ are used to denote sets. And $|\mathcal{M}|$ represents the cardinality of the set $\mathcal{M}$. {\rm Re}$\{\cdot\}$ is the operation to obtain the real part of a complex. $\|\cdot\|_0$ represents the operation of $\ell_0$-norm. $<A,B> = A^HB$.

\section{System Model \label{sec: System Model}}\label{section:2}
In this section, we consider the downlink transmission process with compression transmission strategy in CF MU-MIMO MC system. Multi-users are scheduled in time-frequency resources according to three scheduling methods. Based on the user scheduling model, we formulate the downlink transmission signals received by different users. To this end, the WTP of the CF system both in the FBL regime and the INFBL regime are derived.   
\subsection{Compression Transmission Strategy}
\begin{figure*}[htp]
    \centering
    \includegraphics[width=6.8in]{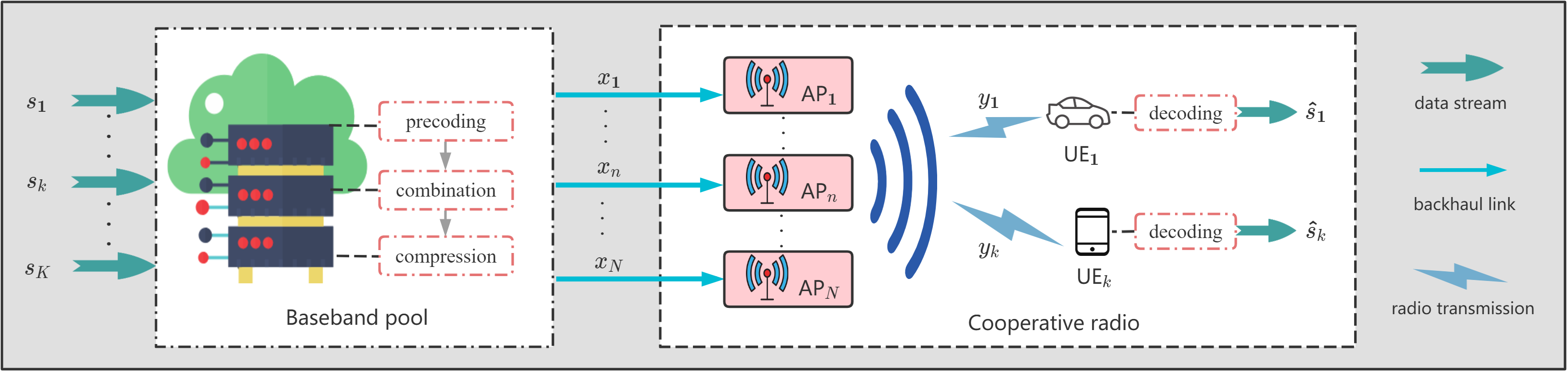}
    \caption{\fontsize{10pt}{\baselineskip}\selectfont System Model of an MU-MIMO Cell-free System.}
    \label{fig_systemModel}
  \end{figure*}

In Fig. \ref{fig_systemModel}, $N$ APs are connected to a BBP in a CF system, hence the channel state informations are shared among all APs and the BBP. Each AP is equipped with $M$ antennas, $K$ single antenna users are distributed in a wide area. 
We denote that $\mathcal{N}=\{1,\cdots,N\}$, $\mathcal{K}=\{1,\cdots,K\}$. Different from the data-sharing transmission strategy that the information precoding is operated in local (i.e., in AP), in compression transmission strategy, information symbols ($s_k, \forall k\in\mathcal{K}$) of users (UE$_k, \forall k\in\mathcal{K}$) are first centrally precoded for different APs by the CPU in the baseband pool. 
After that, the precoded symbols are combined according to their corresponding AP. Hence, we can denote the combined signal for AP$_n, \forall n\in \mathcal{N}$ as follow. 
\begin{equation}
  \label{xn}
  \bm{\overline{x}}_n = \sum\limits_{k=1}^K{\bm{w}_{kn} s_k} ,
\end{equation}
where $\bm{w}_{kn} \in \mathbb{C}^{M\times 1}$ is the beamforming vector of AP$_n$ and UE$_k$. And $s_k$ is the normalized information symbol, i.e., $\mathbb{E}[{|s_k|^2}] = 1$.

Afterward, CPU forwards the compressed signals to their corresponding AP. The received signal in AP$_n$ is expressed as
\begin{equation}
  \label{compressed-signal}
  \bm{x}_n = \bm{\overline{x}}_n + \bm{e}_n,
\end{equation}
where $\bm{e}_n\in \mathbb{C}^{M\times 1}$ is the quantization noise distributed as the Gaussian distribution with variance $\sigma_e^2$, which is introduced by signal compression. 

\subsection{User Scheduling Schemes} 
\begin{figure}[htbp]
  \centering
  \subfigure[RA-SU-MIMO.]{
  \begin{minipage}[t]{0.5\linewidth}\label{fig_su-mimo}
  \flushleft
  \includegraphics[width=1.65in]{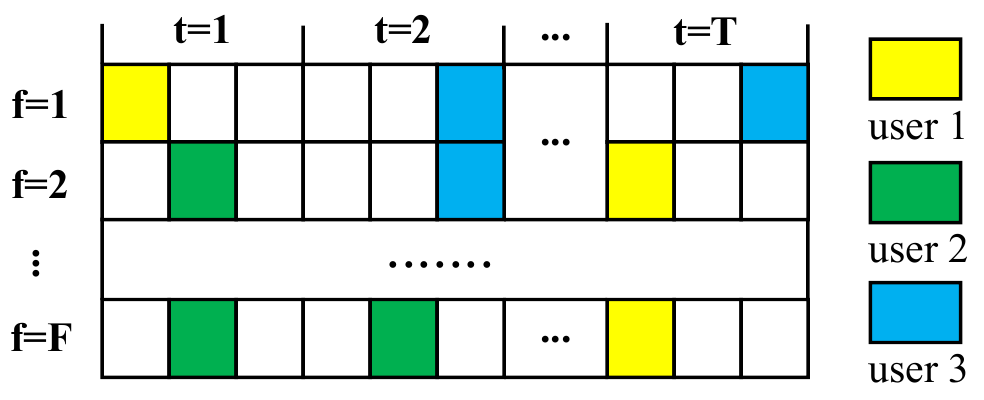}
  \end{minipage}%
  }%
  \subfigure[RA-FU-MIMO.]{
  \begin{minipage}[t]{0.5\linewidth}\label{fig_mu-mimo-all}
  \flushleft
  \includegraphics[width=1.65in]{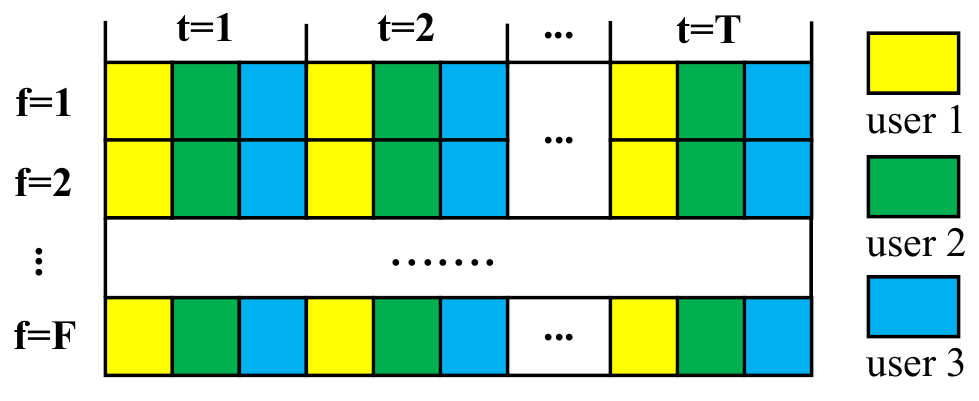}
  \end{minipage}%
  }%

  \subfigure[RA-MU-MIMO.]{
  \begin{minipage}[t]{0.5\textwidth}\label{fig_mu-mimo}
  \flushleft
  \includegraphics[width=3.5in]{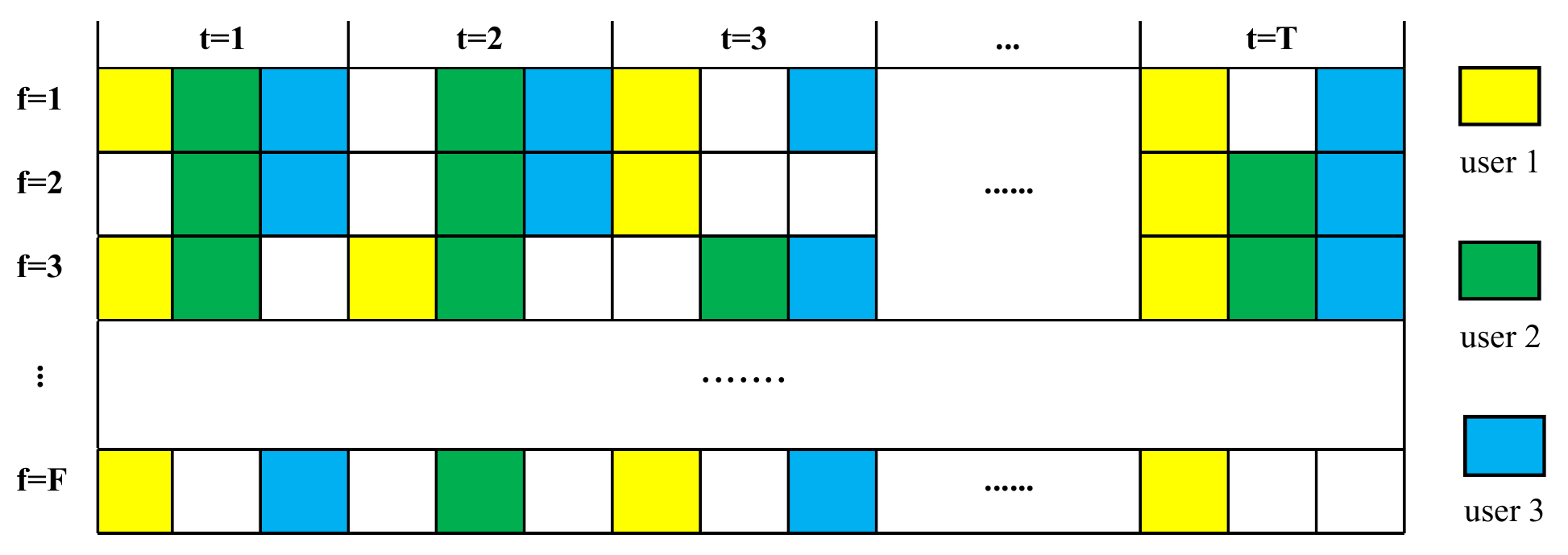}
  \end{minipage}
  }%
  \centering
  \caption{ Different user scheduling schemes in MU-MIMO cell-free systems (a) RA-SU-MIMO: resource allocation in single-user scheduling MIMO scenario; (b) RA-FU-MIMO: resource allocation in full-user scheduling MIMO scenario; (c) RA-MU-MIMO: resource allocation in multi-user scheduling MIMO scenario}
  \label{fig_timeSlotFigure}
\end{figure}

In the downlink signal transmission stage, APs collaboratively transfer the compressed signals to the scheduled users on the time-frequency resource block (RB), which has $F$ subcarriers and $T$ time slots, a total of $T \times F$ resource elements (REs). Three user scheduling schemes are illustrated in Fig. \ref{fig_timeSlotFigure}. The detailed explanation of each user scheduling scheme is listed as 
\begin{itemize}
  \item RA-SU-MIMO (Fig. \ref{fig_su-mimo}: resource allocation of single-user MIMO): up to one user is scheduled in each RE, indicating no interference between users.
  \item RA-FU-MIMO (Fig. \ref{fig_mu-mimo-all}: resource allocation of full-user MIMO): in each RE, all users are scheduled, which means no user scheduling operation at all. 
  \item RA-MU-MIMO (Fig. \ref{fig_mu-mimo}: resource allocation of multi-user MIMO): multiple users are scheduled based on the channel conditions in each RE to maximize the system performance.
\end{itemize}

We point out that RA-MU-MIMO scheme includes the other two schemes. Namely, RA-SU-MIMO and RA-FU-MIMO schemes are two special cases of RA-MU-MIMO scheme.
Therefore, we analyze the downlink signal transmission process based on the RA-MU-MIMO scheme.
In each RE, we denote the transmitting beamforming vector of UE$_k$ as $\bm{w}_k^{tf} = [(\bm{w}_{k1}^{tf})^H,\cdots,(\bm{w}_{kN}^{tf})^H]^H$, $\forall \{t,f\} \in \{\mathcal{T}, \mathcal{F}\}, \mathcal{T}=\{1,\cdots,T\}, \mathcal{F}=\{1,\cdots,F\}$, and $ \bm{w}_{kn}^{tf}\in \mathbb{C}^{M\times 1}, \forall n\in \mathcal{N}$ is the beamforming vector between AP$_n$ and UE$_k$. Hence, the transmitting signal in time slot $t$ and subcarrier $f$ is expressed as
\begin{equation}
  \label{eq_xt}
  \bm{x}^{tf} = \sum_{k\in \mathcal{K}} \bm{w}_k^{tf} s_k^{tf} + \bm{e}^{tf},
\end{equation} 
where $s_k^{tf}$ is the normalized information symbol of UE$_k$ in time slot $t$ and subcarrier $f$, i.e., $\mathbb{E}[{|s_k^{tf}|^2}] = 1$.
$\bm{e}^{tf} = [\bm{e}_1^{tf},\cdots,\bm{e}_N^{tf}]^H$ is the quantization noise vector transmitted from all the $N$ APs. If $\bm{w}_k^{tf} = \bm{0}$ represents UE$_k$ is not scheduled, and vice versa.

In each RE, we denote the channel fading coefficient of UE$_k$ as $\bm {h}_k^{tf} = [(\bm{h}_{k1}^{tf})^H,\cdots,(\bm{h}_{kN}^{tf})^H]^H$, where $\bm{h}_{kn}^{tf} \in \mathbb{C}^{M\times 1}, \forall \{k,n,t,f\} \in \{\mathcal{K}, \mathcal{N}, \mathcal{T}, \mathcal{F}\}$. In the sequal, the received signal of UE$_k$ from APs in time slot $t$ and subcarrier $f$ is denoted as
\begin{equation}
\label{eq_ykt}
\begin{aligned}
y_k^{tf} &= (\bm{h}_k^{tf})^H \sum_{k=1}^K \bm{w}_k^{tf} s_k^{tf} + n_k^{tf}\\
&=\underbrace{(\bm{h}_k^{tf})^H \bm{w}_k^{tf} s_k^{tf}}_{\rm{desired \enspace signal}} + \underbrace{(\bm{h}_k^{tf})^H \sum_{j \neq k} \bm{w}_j^{tf} s_j^{tf}}_{\rm{interference}} \\
&+ \underbrace{(\bm{h}_k^{tf})^H \bm{e}^{tf}}_{\rm{quantization \enspace noise}} +
\underbrace{n_k^{tf}}_{\rm{background \enspace noise}},
\end{aligned}
\end{equation}
where, $n_k^{tf}$ is the complex additive white Gaussian noise with variance $(\sigma_k^{tf})^2$. Except for the background noise, an additional quantization noise is received in each user due to the operation of signal compression in AP.
The signal-to-interference and noise ratio (SINR) in time slot $t$ and subcarrier $f$ is given by
\begin{equation}
\label{eq_gamakt}
\begin{aligned}
  &\quad\gamma_k^{tf}(\bm w^{tf}) \\
&= \frac{|(\bm{h}_k^{tf})^H \bm{w}_k^{tf}|^2}{\sum\limits_{j\neq k} |(\bm{h}_k^{tf})^H \bm{w}_j^{tf}|^2 + {\sum\limits_{n\in\mathcal{N}}||\bm{h}_{kn}^{tf}\sigma_e^{tf}||_2^2} +(\sigma_k^{tf})^2},
\end{aligned}
\end{equation}
where $\bm w^{tf}$ is the collection of $\{\bm{w}_k^{tf}\}, \forall k\in \mathcal{K}, \forall t\in \mathcal{T}, \forall f\in \mathcal{F}$. And the interferences from other users are treated as noise. 

\subsection{Weighted Throughput in Infinite Blocklength Regime}\label{section:2.c}
Based on the SINR in \eqref{eq_gamakt} and the Shannon capacity theory, we obtain the capacity of UE$_k$ in INFBL regime and the WTP of the whole system as follows.
\begin{align}
    &F_k(\bm w) = \sum_{f=1}^F\sum_{t=1}^T{\rm log}_2(1+\gamma_k^{tf}(\bm{w}^{tf})),\label{shannon capacity}\\
    &C(\bm{w}) = \sum\limits_{k=1}^K \rho_k F_k(\bm w),\label{eq_weigtedThoughput_infinite}
\end{align}
where ${\bm w} = [{\bm w}_1^H,\cdots,{\bm w}_K^H]^H$ is the collection of $\{{\bm w}_k\}$, and $\bm{w}_k=[\bm{w}_k^{11},\cdots,\bm{w}_k^{TF}]^H$ is the collection of $\{{\bm w}_k^{tf}\}, \forall \{k,t,f\} \in \{\mathcal{K}, \mathcal{T}, \mathcal{F}\}$. The weighting factor is denoted as $\rho_k$.

\subsection{Weighted Throughput in Finite Blocklength Regime}\label{section:2.d}
Different from the INFBL regime, the total number of bits transmitted to UE$_k$ in a short packet is denoted as \cite{ref_Ghanem2020tcom7184}
\begin{equation}
\label{eq_Fk}
\Phi_k(\bm w) = \sum_{f \in \mathcal{F}}\sum_{t \in \mathcal{T}} {\rm log}_2 \left(1+\gamma_k^{tf}\right)-Q^{-1}(\epsilon )\sqrt{\sum_{f \in \mathcal{F}}\sum_{t \in \mathcal{T}}V_k^{tf}},
\end{equation}
where $V_k^{tf} = 1- \frac{1}{\left(1+\gamma_k^{tf}\right)^2}$ is the channel dispersion in time slot $t$ and subcarrier $f$, and the $Q^{-1}(\cdot)$ represents the inversion of Gaussian $Q$ function with bit error probability $\epsilon$. Ultimately, we obtain the WTP of the whole system as follow.
\begin{equation}
\label{eq_weigtedThoughput}
\begin{aligned}
  U(\bm{w}) &= \sum_{k\in \mathcal{K}} \rho_k \Phi_k(\bm w)\\
  &= \sum_{k\in \mathcal{K}} \rho_k \bigg(F_k(\bm{w})-G_k(\bm{w})\bigg),
\end{aligned} 
\end{equation}
where $F_k(\bm{w})$ is the system capacity in INFBL regime, $G_k(\bm{w})$ is expressed as the following equation.
\begin{equation}
  \label{eq_Gk}
  G_k(\bm{w}) = Q^{-1}(\epsilon )\sqrt{\sum_{f \in \mathcal{F}}\sum_{t \in \mathcal{T}}V_k^{tf}}.
\end{equation}

\section{Solution in infinite blocklength regime}
In this section, we first formulate the optimization problem for maximizing the WTP of the system in INFBL regime. User scheduling algorithm and beamformer optimization algorithm are proposed to address the WTP maximization problem.
\subsection{Problem Formulation}
According to the aforementioned analysis in Section \ref{section:2.c} and Fig. \ref{fig_timeSlotFigure}, we formulate the following maximization problem of the WTP in INFBL regime.
\begin{subequations}
    \label{eq:P_ori_infinite}
  \begin{align}
  \max_{\bm{w}} \quad & C(\bm{w})\\
  \mbox{s.t.}\quad & F_k(\bm{w}) \geq b_k, \forall k\in \mathcal{K}, \label{P_ori_b_in}\\
  & \sum_{k\in \mathcal{K}} \sum_{f\in \mathcal{F}} \|\bm{w}_{k}^{tf} \Vert_2^2  \leq p_t, \forall f\in \mathcal{F}, \forall t\in \mathcal{T}.  \label{P_ori_c_in}
  \end{align}
\end{subequations}

In \eqref{eq:P_ori_infinite}, constraint \eqref{P_ori_b_in} ensures that the minimum number of bits for UE$_k$ is $b_k$.  Constraint \eqref{P_ori_c_in} represents that the total transmitted power on all subcarriers during the $t$-th transmission time slot should be no more than the maximum transmission power of the system.

We further introduce a binary variable $\zeta_k^{tf}\in \{0,1\}, \forall \{k,t,f\} \in \{\mathcal{K}, \mathcal{T}, \mathcal{F}\}$ to represent which user is scheduled in each RE. 
Namely, $\zeta_k^{tf}$ equals 1 means that ${\rm UE}_k$ is scheduled in RE $\{t,f\}$, but not vice versa. Therefore, the expression of SINR in equation \eqref{eq_gamakt} can be rewritten as
\begin{equation}
  \label{eq_gamakt_d}
  \begin{aligned}
    &\quad\gamma_k^{tf}(\bm w^{tf}, \bm \zeta^{tf}) \\
    &= \frac{|(\bm{h}_k^{tf})^H \zeta_k^{tf}\bm{w}_k^{tf}|^2}{\sum\limits_{j\neq k} |(\bm{h}_k^{tf})^H \zeta_j^{tf}\bm{w}_j^{tf}|^2 + {\sum\limits_{n\in\mathcal{N}}||\bm{h}_{kn}^{tf}\sigma_e^{tf}||_2^2} +(\sigma_k^{tf})^2},
  \end{aligned}
  \end{equation}
where $\bm{\zeta}^{tf}=[\zeta_1^{tf},\cdots,\zeta_K^{tf}]$. 

In the sequal, we transform problem \eqref{eq:P_ori_infinite} into problem \eqref{eq:P_ori_infinite_d} 
\begin{subequations}
    \label{eq:P_ori_infinite_d}
  \begin{align}
  \max_{\bm{w},\bm{\zeta}} \quad & C(\bm{w},\bm{\zeta})\\
  \mbox{s.t.}\quad & F_k(\bm{w},\bm{\zeta}) \geq b_k,  \label{P_ori_c_in_c}\\
  &\sum_{k\in \mathcal{K}} \sum_{f\in \mathcal{F}} \|\zeta_k^{tf}\bm{w}_{k}^{tf} \Vert_2^2  \leq p_t,  \label{P_ori_c_in_d}\\
  &\zeta_k^{tf} \in \{0,1\}, \forall k\in \mathcal{K},\forall t\in \mathcal{T},\forall f\in \mathcal{F}, \label{P_ori_c_in_e}
  \end{align}
\end{subequations}
where $\bm \zeta$ is the collection of $\{\bm{\zeta}_k\}$, and $\bm{\zeta}_k=[\zeta_k^{11},\cdots,\zeta_k^{TF}],\forall k\in \mathcal{K}$.

The objective function of problem \eqref{eq:P_ori_infinite_d} is non-convex. And it is a mixed integer nonlinear programming problem \cite{MINIPproblem}  due to the binary variable $\bm{\zeta}$ \eqref{P_ori_c_in_e} and the non-convex constraint \eqref{P_ori_c_in_c}, which is hard to be tackled. To solve this NP-hard problem, we utilize the alternating iterative optimization algorithm. In each iteration, the user scheduling algorithm based on the gene-aided (GA) algorithm \cite{GAalgorithm} is first proposed to optimize the binary variable $\bm{\zeta}$ with fixed beamforming vector $\bm{w}$ obtained from the last iteration. Afterward, a beamforming optimization algorithm is proposed to further optimize the system performance with the fixed user scheduling scheme obtained before. Alternating iterative optimization stops until the algorithm converges.

\subsection{User Scheduling}
\begin{figure}[htp]
  \centering
  \includegraphics[width=3.5in]{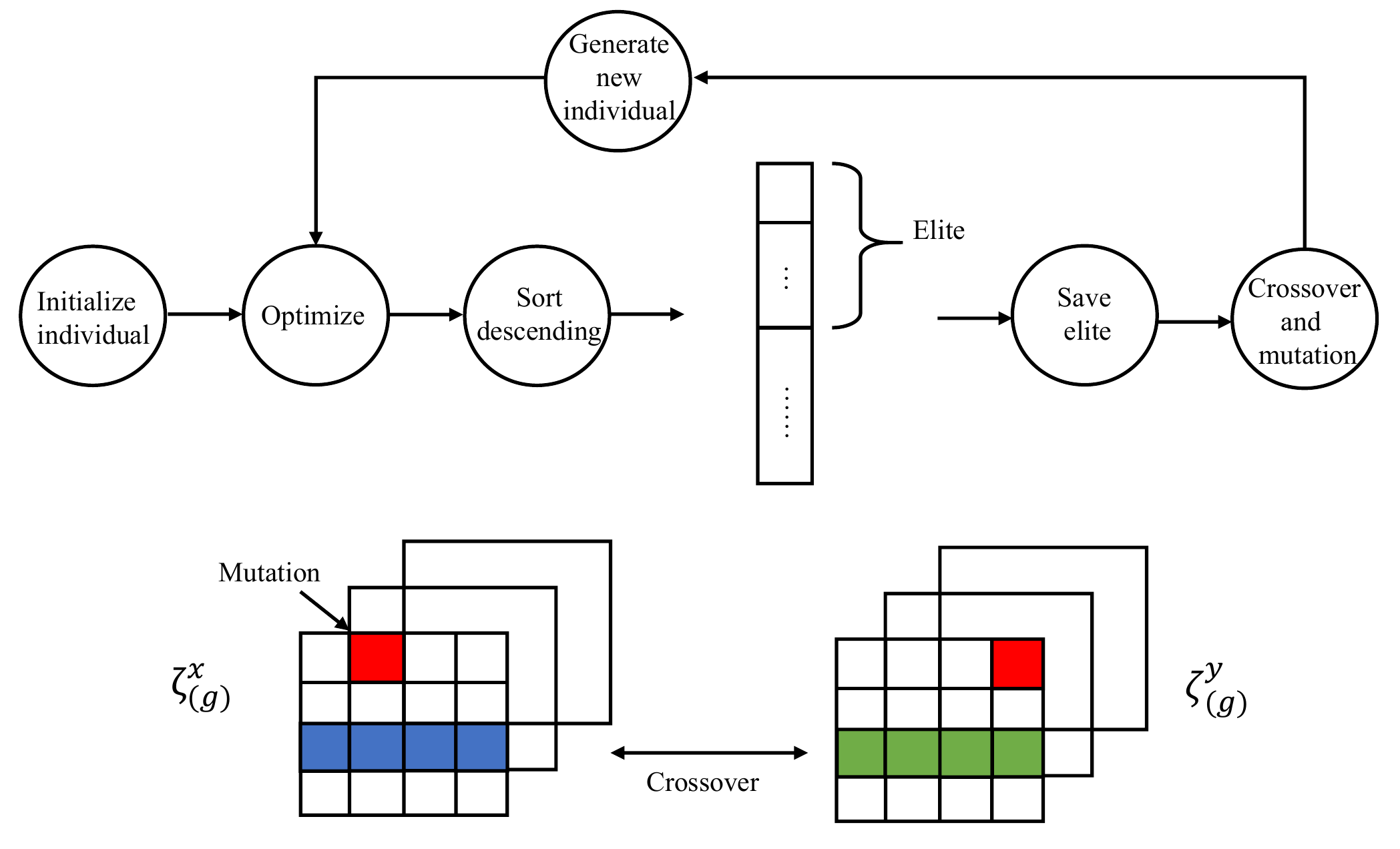}
  \caption{\fontsize{10pt}{\baselineskip}\selectfont Process of genetic selection.}
  \label{fig_ga}
\end{figure}

In this subsection, the user scheduling scheme based on the gene-aided algorithm is used to solve the binary integer programming (BIP) problem. Fig. \ref{fig_ga} describes the process of user scheduling from the selection of variable $\bm \zeta$. The detailed explanations are listed as follows. 

\begin{itemize}
  \item \textbf{Initialize individual}: since the variable $\bm{\zeta}$ is a binary variable which satisfies the coding space of genetic algorithm, we generate individual with binomial distribution and map the individual to $\bm{\zeta}$ directly. Namely, $\bm{\zeta}_{(g)}=[\bm{\zeta}_{(g)}^1,\cdots,\bm{\zeta}_{(g)}^{P_{op}}]$ at the $g$-th generation.
  \item \textbf{Optimize}: after generating the individual, i.e., the time-frequency resource $\bm{\zeta}$ is fixed, we next calculate the fitness of each individual $x$. Namely, calculate the WTP $U_{(g)}^x(\bm{\zeta}_{(g)}^{x})$ according to the individual and the beamforming design result (Algorithm \ref{alg:algGWMMSE} and \ref{alg:algSCA}).
  \item \textbf{Sort descending}: sort the fitnesses in descending order. Hence, the first few fitnesses have the largest values $[U_{(g)}^1,\cdots,U_{(g)}^{e}]$. 
  \item \textbf{Save elite}: select and keep the first $e$ individuals $[\bm{\widetilde{\zeta}}_{(g)}^1,\cdots,\bm{\widetilde{\zeta}}_{(g)}^e]$ to the next generation.
  \item \textbf{Crossover and mutation}: 1). the crossover probability $p_{c(g)}$ is generated randomly, if $p_{c(g)} \geq p_c$, crossover happens between individuals, e.g., $\bm{\zeta}_{(g)}^{x} \Leftrightarrow \bm{\zeta}_{(g)}^{y}$ means that the randomly selected row of $\bm{\zeta}_{(g)}^{x}$ (the blue row) and the corresponding green row of $\bm{\zeta}_{(g)}^{y}$ are exchanged between individuals $x$ and $y$. 2). the mutation probability $p_{m(g)}$ is generated randomly, if $p_{m(g)} \geq p_m$, mutation occurs in each individual itself, e.g., the red element of $\bm{\zeta}_{(g)}^{y}$ changes. 
  \item \textbf{Generate new individual}: crossover and mutation individuals are involved in the next round of genetic selection as the new population. 
\end{itemize}  

Algorithm \ref{alg:algGA} summarizes the process of iterative optimization with user scheduling and beamforming design.
\begin{algorithm}
    \caption{{Joint user scheduling and beamforming design algorithm (USBDA)}}
    \label{alg:algGA}
    \begin{algorithmic}[1]
    \State{\textbf{Input:}} population size $P_{op}$, elite number $e$, rate $r$, crossover probability $p_c$, mutation probability $p_m$, maximum generation algebra $g_{\text{max}}$.
    \State{\textbf{Initialize:}} $g=0$, population set $\bm{\zeta}_{(g)} = [\bm{\zeta}_{(g)}^1,\cdots,\bm{\zeta}_{(g)}^{P_{op}}]$.
    \State \textbf{while} $g \textless g_{\text{max}}$
    \begin{enumerate}
        \State Calculate fitness $U_{(g)}^x$ of each individual $x$ according to beamforming design algorithm (i.e., MMSE beamformer/Algorithm \ref{alg:algGWMMSE}/Algorithm \ref{alg:algSCA}).
        \State Sort fitness $[U_{(g)}^1,\cdots,U_{(g)}^{P_{op}}]$ in descending order.
        \State Select the first $e$ individuals $[\bm{\widetilde{\zeta}}_{(g)}^1,\cdots,\bm{\widetilde{\zeta}}_{(g)}^e]$ as the elites are reserved for the next generation. 
        \State{\textbf{if} $p_{c(g)} \geq p_c$} (crossover)
         \begin{enumerate}
             \State calculate the weight of the fitnesses
             \State exchange the corresponding row between individuals.
         \end{enumerate}
         \State{\textbf{end if}}
         \State{\textbf{if} $p_{m(g)} \geq p_m$} (mutation)
         \begin{enumerate}
             \State randomly select row number
             \State mutation goes in the direction of the largest weight.
         \end{enumerate}
         \State{\textbf{end if}}
         \State obtain the new population.
         \State{\textbf{if} achieve the maximum generation number $g_{\rm max}$}
         \begin{enumerate}
             \State{\textbf{break}}
         \end{enumerate}
         \State{\textbf{end if}}
    \end{enumerate}
    \State \textbf{end for}
    \State{\textbf{Output:} $\bm{\zeta}$}
    \end{algorithmic}
\end{algorithm}

\subsection{Beamforming Design in Infinite Blocklength Regime}\label{section:3.c}

With fixed user scheduling scheme, we aim to maximize the WTP of the system by beamforming design. To simplify the notation, we denote $\bm \chi_k^{tf} =\zeta_k^{tf} \bm{w}_k^{tf}$. Therefore, we obtain the beamforming design problem as the following equation.
\begin{subequations}
    \label{eq:P_ori_infinite_bf}
  \begin{align}
  \max_{\bm{\chi}} \quad & C(\bm{\chi})\\
  \mbox{s.t.}\quad& F_k(\bm{\chi}) \geq b_k, \forall k\in \mathcal{K},\label{P_ori_c_in_f} \\
  &\sum_{k\in \mathcal{K}} \sum_{f\in \mathcal{F}} \|\bm{\chi}_{k}^{tf} \Vert_2^2  \leq p_t, \forall t\in \mathcal{T},  \label{P_ori_c_in_bf}
  \end{align}
\end{subequations}
where $\bm \chi$ is the collection of $\{\bm{\chi}_k^{tf}\}, \forall k\in \mathcal{K}, \forall t\in \mathcal{T}, \forall f\in \mathcal{F}$, and $\bm{\chi}_k^{tf} \in \mathbb{C}^{M\times 1}$.

It is still challenging to solve problem \eqref{eq:P_ori_infinite_bf} due to the non-convex objective function and constraint \eqref{P_ori_c_in_f}. 
To solve this problem, we first denote $\mathbf{X}_k^{tf} = \bm{\chi}_k^{tf}(\bm{\chi}_k^{tf})^H$, $\mathbf{H}_k^{tf} = \bm{h}_k^{tf} (\bm{h}_k^{tf})^H$. Hence, SINR \eqref{eq_gamakt_d} and throughput \eqref{shannon capacity} of UE$_k$ are transformed into the following equations.
\begin{equation}\label{eq_gama_sdr}
    \begin{aligned}
    \gamma_k^{tf}(\mathbf{X})=\frac{{\rm Tr}(\mathbf{H}_k^{tf} \mathbf{X}_k^{tf})}{\sum\limits_{j\neq k} {\rm Tr}(\mathbf{H}_k^{tf} \mathbf{X}_j^{tf}) + (\sigma_{ke}^{tf})^2},
  \end{aligned}
\end{equation}
\begin{equation}
    F_k(\mathbf{X}) = \sum_{t\in \mathcal{T}}\sum_{f\in \mathcal{F}} {\rm log_2} (1+\gamma_k^{tf}(\mathbf{X})), \label{sc_X}
\end{equation}
where $(\sigma_{ke}^{tf})^2 = {(\sigma_e^{tf})^2{{\rm Tr}(\mathbf{H}_k^{tf})}} +(\sigma_k^{tf})^2$ is the summation of quantization noise and background noise at ${\rm UE}_k$ in time slot $t$ and subcarrier $f$.
 $\mathbf{X}$ is the collection of $\mathbf{X}_k^{tf}, \forall k\in \mathcal{K}, \forall t\in \mathcal{T}, \forall f\in \mathcal{F}$.

For clarity, we define the following expressions.
\begin{align}
  &\Upsilon(\mathbf{X})\triangleq {\rm log_2}\big(1+\gamma_k^{tf}(\mathbf{X})\big). \label{eq_upsilon}\\
  &\Upsilon_1(\mathbf{X})\triangleq \sum_{k\in \mathcal{K}}{\rm Tr}(\mathbf{H}_k^{tf} \mathbf{X}_k^{tf}) + (\sigma_k^{tf})^2.\label{eq_Upsilon_1}\\
  &\Upsilon_2(\mathbf{X})\triangleq \sum_{j\neq k}{\rm Tr}(\mathbf{H}_k^{tf} \mathbf{X}_j^{tf}) + (\sigma_k^{tf})^2. \label{eq_upsilon_2}
\end{align}

Through some mathematical transformations after substituting \eqref{eq_gama_sdr} into \eqref{eq_upsilon}, we can rewrite equation \eqref{eq_upsilon} and \eqref{sc_X} as
\begin{equation}
  \Upsilon(\mathbf{X})={\rm log}_2\big(\Upsilon_1(\mathbf{X})\big)-{\rm log}_2\big(\Upsilon_2(\mathbf{X})\big). \label{eq_upsilonLog}
\end{equation}

Equation \eqref{eq_upsilonLog} is still a non-convex function. Nevertheless, it has the form of difference of convex (DC), which can be approximated to a convex one by first order Taylor approximation. To this end, we first calculate the first derivative of \eqref{eq_upsilon_2} with respect to $\mathbf{X}_j^{tf}$ as follow.
\begin{equation}
  \frac{\partial \Upsilon_2(\mathbf{X})}{\partial \mathbf{X}_j^{tf}} = \frac{\mathbf{H}_k^{tf}}{\Upsilon_2(\mathbf{X}){\rm ln}(2)}.
\end{equation}

Therefore, equation \eqref{eq_upsilonLog} is transformed into the convex one as shown in equation \eqref{eq_upsilonLogApp}.
\begin{equation}\label{eq_upsilonLogApp}
  \begin{aligned}
  &\quad\overline{\Upsilon}(\mathbf{X})\\
  &={\rm log}_2\big(\Upsilon_1(\mathbf{X})\big)-\sum_{j\neq k}<\frac{\partial \Upsilon_2(\mathbf{X})}{\partial \mathbf{\hat{X}}_j^{tf}},\big(\mathbf{X}_j^{tf}-\mathbf{\hat{X}}_j^{tf}\big)>\\
  &={\rm log}_2\big(\Upsilon_1(\mathbf{X})\big)-\sum_{j\neq k}{\rm Tr}\bigg(\frac{\partial \Upsilon_2(\mathbf{X})}{\partial \mathbf{\hat{X}}_j^{tf}}\big(\mathbf{X}_j^{tf}-\mathbf{\hat{X}}_j^{tf}\big)^H\bigg).
  \end{aligned}
\end{equation}
where $\mathbf{\hat{X}}_j^{tf}$ is obtained from the last iteration result.

For notation simplicity, we define $\overline{F}_k(\mathbf{X})\triangleq \sum_{t\in \mathcal{T}}\sum_{f\in \mathcal{F}} \overline{\Upsilon}(\mathbf{X})$. In the sequal, the equivalence of problem \eqref{eq:P_ori_infinite_bf} is denoted as
\begin{subequations}
    \label{eq:P_ori_infinite_sd}
  \begin{align}
  \max_{\mathbf{X}} \quad & \sum_{k\in \mathcal{K}}\overline{F}_k(\mathbf{X})\\
  \mbox{s.t.}\quad& \overline{F}_k(\mathbf{X}) \geq b_k, \label{P_ori_c_in_f_sd} \\
  &\sum_{k\in \mathcal{K}} \sum_{f\in \mathcal{F}} {\rm Tr}(\mathbf{X}_k^{tf}) \leq p_t,   \label{P_ori_c_in_bf_sd}\\
  &{\rm Rank}(\mathbf{X}_k^{tf}) \leq 1, \label{rank}\\
  & \mathbf{X}_k^{tf} \succeq 0, \forall k\in \mathcal{K}, \forall t\in \mathcal{T}, \forall f\in \mathcal{F}, \label{sdrCon}
  \end{align}
\end{subequations}
where Rank($\cdot$) represents the rank of a matrix and $\succeq$ stands for the semi-positive definite.

Since the constraint \eqref{rank} is non-convex, we first drop it by utilizing the semidefinite relaxation (SDR) method, thus problem \eqref{eq:P_ori_infinite_sd} is transformed into problem \eqref{eq:P_ori_infinite_sdr}.
\begin{subequations}\label{eq:P_ori_infinite_sdr}
  \begin{align}
  \max_{\mathbf{X}} \quad & \sum_{k\in \mathcal{K}}\overline{F}_k(\mathbf{X})\\
  \mbox{s.t.}\quad& \eqref{P_ori_c_in_f_sd}, \eqref{P_ori_c_in_bf_sd}, \eqref{sdrCon}. 
  \end{align}
\end{subequations}

It is not difficult to see that problem \eqref{eq:P_ori_infinite_sdr} is convex relative to variable $\mathbf{X}$, which can be solved by some standard optimization tools (e.g., CVX). Furthermore, the  eigendecomposition method or Gaussian randomization method is adopted to recover the BF vector $\bm{\chi}$ from $\mathbf{X}$.  

Algorithm \ref{alg:algGWMMSE} concludes the beamforming design based on the SDR method in INFBL regime.
\begin{algorithm}

  \caption{{Beamforming design algorithm in INFBL regime (BF-INFBL)}}
  \label{alg:algGWMMSE}
  \begin{algorithmic}[1]
  \State{\textbf{Input:}} the channel gain $\bm{h}_k^{tf}, \forall k,\in \mathcal{K}, \forall t\in \mathcal{T}, \forall f\in \mathcal{F}$, maximum iteration number is $i_{\text{max}}$.
  \State{\textbf{Initialize:}} the beamforming vector $(\bm{\chi}_t^{tf})^{(0)}$ is initialized by MMSE beamformer. Calculate $(\mathbf{X}_k^{tf})^{(0)} = (\bm{\chi}_t^{tf})^{(0)}((\bm{\chi}_t^{tf})^{(0)})^H$, $\mathbf{H}_k^{tf} = \bm{h}_k^{tf} (\bm{h}_k^{tf})^H$. 
  \State \textbf{for} each iteration $i$
  \begin{enumerate}
      \State Calculate $\gamma_k^{tf}((\mathbf{X})^{(i)})$ according to equation \eqref{eq_gama_sdr}.
      \State Calculate $\overline{\Upsilon}((\mathbf{X})^{(i)})$ according to \eqref{eq_upsilonLogApp}.
      \State Solve problem \eqref{eq:P_ori_infinite_sdr} to obtain $(\mathbf{X})^{(i)}$.
      \State Calculate the WTP according to the objective function of \eqref{eq:P_ori_infinite_sdr}.
      \State Update the iteration number: $i = i + 1$.
      \begin{enumerate}
        \State $ (\mathbf{X}_k^{tf})^{(i+1)} \gets (\mathbf{X}_k^{tf})^{(i)}$.
        \State $ i \gets i + 1 $.
      \end{enumerate}
       \State{\textbf{if} achieve the maximum iteration number $i_{\rm max}$}
       \begin{enumerate}
           \State{\textbf{break}}.
       \end{enumerate}
       \State{\textbf{end if}}
  \end{enumerate}
  \State \textbf{end for}
  \State{\textbf{Output:} $(\mathbf{X})^{i_{max}}$}.
  \State Recovery the beamforming vector $\bm{\chi}$ from $\mathbf{X}$ by eigendecomposition method or Gaussian randomization.
  \State Calculate the WTP according to the objective function of problem \eqref{eq:P_ori_infinite_sdr}.
  \end{algorithmic}
\end{algorithm}

\section{Solution in finite blocklength regime}

\subsection{Problem Formulation}
In this section, we formulate and address the optimization problems for maximizing the WTP of the whole system by user scheduling and beamforming design in FBL regime.
As shown in Fig. \ref{fig_timeSlotFigure}, the colored patch indicates that the corresponding user is scheduled while the blank patch indicates no service. Hence, the latency in FBL regime is equal to or greater than the blocklength. 

Based on Fig. \ref{fig_timeSlotFigure}, we formulate the following optimization problem to maximize the WTP of the whole system in FBL regime.
\begin{subequations}\label{eq:P_ori}
  \begin{align}
  \max_{\bm{w}} \quad & U(\bm{w})\\
  \mbox{s.t.}\quad &F_k(\bm{w})-G_k(\bm{w}) \geq b_k, \forall k\in \mathcal{K}, & \label{P_ori_b}\\
  & \sum_{k\in \mathcal{K}} \sum_{f\in \mathcal{F}} \|\bm{w}_{k}^{tf} \Vert_2^2  \leq p_t, \forall t\in \mathcal{T}, \forall f\in \mathcal{F}. \label{P_ori_c}
  \end{align}
\end{subequations}

In \eqref{eq:P_ori}, constraint \eqref{P_ori_b} represents the minumum transmit data requirement to satisfy the QoS of each user. 
Constraint \eqref{P_ori_c} means that the total transmission power on the total subcarriers at $t$-th transmission duration should be less than the total transmission power. Similar to the user scheduling scheme in INFBL regime, we rewrite problem \eqref{eq:P_ori} into the following equivalent problem by introducing a binary variable $\zeta$.
\begin{subequations}\label{eq:P_ori_d}
  \begin{align}
  \max_{\bm{w},\bm{\zeta}} \quad & U(\bm{w},\bm{\zeta})\\
  \mbox{s.t.}\quad &F_k(\bm{w},\bm{\zeta})-G_k(\bm{w},\bm{\zeta}) \geq b_k,  \label{P_ori_b_d} \\
  & \sum_{k\in \mathcal{K}} \sum_{f\in \mathcal{F}} \|\zeta_k^{tf}\bm{w}_{k}^{tf} \Vert_2^2  \leq p_t,  \label{P_ori_c_d}\\
  & \zeta_k^{tf} \in \{0,1\}, \forall k\in \mathcal{K}, \forall t\in \mathcal{T}, \forall f\in \mathcal{F}. \label{P_ori_e_d}
  \end{align}
\end{subequations}
\newtheorem{remark}{Remark}
\begin{remark}
  Similarly, $\zeta_k^{tf}$ denotes whether UE$_k$ is scheduled in resource element $\{t,f\}$. In addition, we point out that the summation of time slots in each subcarrier represents the maximum latency requirement in FBL regime. And the relationship between the blocklength $L_k^f$ of UE$_k$ in subcarrier $f$ and the indicator factor $\bm{\zeta}_k^f$ is fomulated as the $\ell_0$-norm of $\bm{\zeta}_k^f$, i.e.,
  \begin{equation}
    L_k^f = \|\bm{\zeta}_k^f\|_0, \forall k\in \mathcal{K}, \forall f\in \mathcal{F}.
  \end{equation}

Furthermore, the transmission latency is greater than or equal to the blocklength but not more than the maximum transmission latency $T$ to meet the latency requirement for URLLC transmission.                  

\end{remark}

\subsection{Beamforming Design in Finite Blocklength Regime}
As same as problem \eqref{eq:P_ori_infinite_d}, problem \eqref{eq:P_ori_d} is an MINLP problem which is intractable to address. Hence, we decompose the problem into two subproblems, including user scheduling and beamforming design. And these two subproblems are solved in turn iteratively. Since the user scheduling algorithm has been given in Algorithm \ref{alg:algGA}, here we omit it due to space limitations. 

In the sequel, we aim to optimize the beamforming vector $\bm{w}_k^{tf}$ with fixed $\bm{\zeta}$. As same as Section \ref{section:3.c}, by utilizing the semidefinite relaxation method, problem \eqref{eq:P_ori_d} is transformed as 
\begin{subequations}\label{eq:P_ori_two}
  \begin{align}
  \max_{{\mathbf{X}}} \quad & U(\mathbf{X})\\
  \mbox{s.t.}\quad & F_k(\mathbf{X})-G_k(\mathbf{X}) \geq b_k,\label{P_ori_two_b}\\
  & \sum_{k\in \mathcal{K}} \sum_{f\in \mathcal{F}} {\rm Tr}\left(\mathbf{X}_k^{tf}\right)  \leq p_t, \label{P_ori_two_c}\\ 
  & {\rm Rank}(\mathbf{X}_k^{tf}) \leq 1,\label{P_ori_two_f}\\
  &  \mathbf{X}_k^{tf} \succeq 0, \forall k\in \mathcal{K}, \forall f\in \mathcal{F}, \forall t\in \mathcal{T}. \label{P_ori_two_g}
  \end{align}
\end{subequations}

For simplicity, we let
\begin{equation}\label{interf}
  \begin{aligned}
    I_k^{tf}(\mathbf{X}) \triangleq  \sum_{j\neq k}{\rm Tr}\left(\mathbf{H}_k^f \mathbf{X}_j^{tf}\right).
  \end{aligned}
\end{equation}

Since the constraint \eqref{P_ori_two_f} is nonconvex, we first drop it. In addition, the objective function is nonconvex, and we introduce a series of auxilary variables $\bm{z}=\{\bm{z}_k, \forall k\in \mathcal{K}$\}, $\bm{z}_k =[z_k^{11},\cdots,z_k^{TF}]^T$ to bound the SINRs, hence the problem \eqref{eq:P_ori_two} is rewritten as
\begin{subequations}\label{eq:P_ori_three}
  \begin{align}
  \max_{\mathbf{X},\bm{z}} \quad & U(\bm{z})\\
  \mbox{s.t.}\quad &F_k(\bm{z}_k)-G_k(\bm{z}_k) \geq b_k, \label{P_ori_three_bb}\\ 
  & z_k^0 \leq z_k^{tf} \leq \gamma_k^{tf}, \forall k\in \mathcal{K}, \forall f\in \mathcal{F}, \forall t\in \mathcal{T}, \label{P_ori_three_d}\\
  & \eqref{P_ori_two_c},\eqref{P_ori_two_g}, \label{P_ori_three_c}
  \end{align}
\end{subequations}
where $z_k^0$ is the minimum SINR boundary, $U(\bm{z}_k)$, $F_k(\bm{z}_k)$, $G_k(\bm{z}_k)$ are denoted as
\begin{align}
  &U(\bm{z}) = \sum_{k\in \mathcal{K}} \rho_k \bigg{(}F_k(\bm{z}_k)-G_k(\bm{z}_k)\bigg{)}.\label{eq_U_zk}\\
  &F_k(\bm{z}_k) = \sum_{f\in \mathcal{F}} \sum_{t\in \mathcal{T}} {\rm log}_2 \left(1+\bm{z}_k^{tf}\right).\\
  &G_k(\bm{z}_k) = Q^{-1}(\epsilon )\sqrt{V_k(\bm{z}_k)}.
\end{align}
\begin{align}
  &V_k(\bm{z}_k) = \sum_{f\in \mathcal{F}} \sum_{t\in \mathcal{T}}\left(1-(1+z_k^{tf})^{-2}\right).
\end{align}

Since the objective function in \eqref{eq_U_zk} has the form of difference of convex, we can approximate $G_k\left(\bm{z}_k\right)$ by first order Taylor expansion.
\begin{equation}
  \label{GkTaylor}
  \begin{aligned}
  &\quad G_k\left(\bm{z}_k\right) \leq \overline{G}_k\left(\bm{z}_k\right)\\
  &=G_k\left(\bm{z}_k^{(iter)}\right) + \bigtriangledown_{z_k}G_k\left(\bm{z}_k\right)^T\left(\bm{z}_k-\bm{z}_k^{(iter)}\right), 
  \end{aligned}
\end{equation}
where,
\begin{equation}
  \bigtriangledown_{\bm{z}_k}G_k\left(\bm{z}_k\right) = \frac{1}{\sqrt{V_k\left(\bm{z}_k^{({\rm iter})}\right)}}
    \left[\begin{array}
    {c}
    \left(1+z_k^1\right)^{-3} \\
    \left(1+z_k^2\right)^{-3}\\
    \vdots\\
    \left(1+z_k^T\right)^{-3}
    \end{array}\right].
\end{equation}

To step further, the auxilary variables $\pi_k^{tf}, \theta_k^{tf}, \forall k,t,f$ are introduced to approximate the SINRs, hence the constraint \eqref{P_ori_three_d} can be expressed as the following inequality, 
\begin{equation}\label{sinr_app}
  z_k^0 \leq z_k^{tf} \leq \frac{(\pi_k^{tf})^2}{\theta_k^{tf}} \leq \gamma_k^{tf},
\end{equation}
where 
\begin{equation}
      (\pi_k^{tf})^2 \leq d_k^{tf}{\rm Tr}\left(\mathbf{H}_k^f \mathbf{X}_k^{tf}\right).\label{eq_weigtedThoughput_zk_1}
 \end{equation}
 \begin{equation}
      \theta_k^{tf} \geq I_k^{tf} + (\sigma_{ke}^{tf})^2.\label{eq_weigtedThoughput_zk_2}
 \end{equation}

Since the constraints \eqref{sinr_app} and \eqref{eq_weigtedThoughput_zk_1} are still nonconvex, we further approximate them into the convex one by applying first order Taylor expression which are denoted as the following expressions.
\begin{equation}
  \label{eq_pi_Taylor}
   (\pi_k^{tf})^2 \geq  2\hat{\pi}_k^{tf} \pi_k^{tf} - (\hat{\pi}_k^{tf})^2 .
  \end{equation}
  \begin{equation}
  \label{eq_pi_omega_Taylor}
  \frac{(\pi_k^{tf})^2}{\theta_k^{tf}} \geq \frac{2\hat{\pi}_k^{tf}}{\hat{\theta}_k^{tf}} \pi_k^{tf} - (\frac{\hat{\pi}_k^{tf}}{\hat{\theta}_k^{tf}})^2 \theta_k^{tf} .
  \end{equation}

In the end, we obtain the following approximation problem \eqref{eq:P_ori_four}, which is a convex problem and can be solved by standard optimization tools. The process of beamforming design is summarized in Algorithm \ref{alg:algSCA}.
\begin{subequations}\label{eq:P_ori_four}
  \begin{align}
  \max_{\mathbf{X},\bm{z},\bm{\pi},\bm{\theta}} \quad & \sum_{k=1}^K \rho_k \bigg{(}F_k(\bm{z}_k)-\overline{G}_k(\bm{z}_k)\bigg{)}\\
  \mbox{s.t.}\quad & \eqref{P_ori_three_bb}, \eqref{P_ori_three_c},\eqref{eq_weigtedThoughput_zk_2}\sim\eqref{eq_pi_omega_Taylor}, \label{P_ori_four_d}
  \end{align}
\end{subequations}
where $\bm\pi=\{\pi_k^{tf}, \forall k\in \mathcal{K}, \forall f\in \mathcal{F}, \forall t\in \mathcal{T}\}$, $\bm\theta=\{\theta_k^{tf}, \forall k\in \mathcal{K}, \forall f\in \mathcal{F}, \forall t\in \mathcal{T}\}$.
\begin{algorithm}
  \caption{{Beamforming design in FBL regime (BF-FBL)}}
  \label{alg:algSCA}
  \begin{algorithmic}[1]
  \State{\textbf{Input:}} the channel gain $\bm{h}_k^{tf}, \forall k\in \mathcal{K}, \forall f\in \mathcal{F}, \forall t\in \mathcal{T}$, maximum iteration number is $i_{\text{max}}$.
  \State{\textbf{Initialize:}} the beamforming vector $(\bm{\chi}_k^{tf})^0$ is initialized by MMSE beamformer. In the sequal, $(\mathbf{X}_k^{tf})^{(0)}$, $(z_k^{tf})^{(0)}$, $(\pi_k^{tf})^{(0)}$ and $(\theta_k^{tf})^{(0)}$ are obtained.
  \State \textbf{for} each iteration $i$
  \begin{enumerate}
      \State Solve problem \eqref{eq:P_ori_four} using CVX tool with SDP solver, and obtain the variables of $(\mathbf{X}_k^{tf})^{(i)}$, $(z_k^{tf})^{(i)}$, $(\pi_k^{tf})^{(i)}$ , $(\theta_k^{tf})^{(i)}$.
      \State Calculate the WTP $U(\mathbf{X})$.
      \State Update the following variables 
      \begin{enumerate}
          \State $ (\mathbf{X}_k^{tf})^{(i+1)} \gets (\mathbf{X}_k^{tf})^{(i)}$.
          \State $ (z_k^{tf})^{(i+1)} \gets (z_k^{tf})^{(i)}$.
          \State $ (\pi_k^{tf})^{(i+1)} \gets (\pi_k^{tf})^{(i)}$.
          \State $ (\theta_k^{tf})^{(i+1)} \gets (\theta_k^{tf})^{(i)}$.
          \State $ i = i + 1 $.
      \end{enumerate}
       \State{\textbf{if} achieve the maximum iteration number $i_{\rm max}$}
       \begin{enumerate}
           \State{\textbf{break}}.
       \end{enumerate}
       \State{\textbf{end if}}
  \end{enumerate}
  \State \textbf{end for}
  \State{\textbf{Output:} $(\mathbf{X}_k^{tf})^{(i)}$, $(z_k^{tf})^{(i)}$, $(\pi_k^{tf})^{(i)}$ , $(\theta_k^{tf})^{(i)}$}.
  \State Recovery the beamforming vector $\bm{\chi}_k^{tf}$ from $\mathbf{X}_k^{tf}$ by eigendecomposition method or Gaussian randomization.
  \end{algorithmic}
\end{algorithm}

\subsection{Complexity Analysis}

Since problem \eqref{eq:P_ori_infinite_sdr} consists of $K$ variables, $KTF+K+T$ constraints are solved by the CVX solver within the polynomial time. Therefore, the computational complexity of Algorithm \ref{alg:algGWMMSE} is $\mathcal{O} (K^3(KTF+T+K))$\cite{ref_complextiy}.
Similarly, $4K$ variables and $3KTF+T+K$ constraints in problem \eqref{eq:P_ori_four}. The computational complexity of Algorithm \ref{alg:algSCA} is $\mathcal{O} ((4K)^3(3KTF+T+K))$.
In addition, the computational complexity of the GA method grows linearly with the number of generations and the population increase. 
To this end, the computational complexity of Algorithm \ref{alg:algGA} with Algorithm \ref{alg:algGWMMSE} and Algorithm \ref{alg:algSCA} equal to $\mathcal{O}((P_{op}e)K^3(KTF+T+K))$ and $\mathcal{O}((P_{op}e)(4K)^3(3KTF+T+K))$.

\subsection{Cell-free System vs Centralized Antenna System}

Actually, our proposed user scheduling method can be easily extended to the centralized antenna system (CAS). The transmitting signal in each TF resource from centralized base station (BS) is expressed as
\begin{equation}
  \bm{\hat{x}}^{tf} = \sum\limits_{k\in \mathcal{K}}\bm{\hat{w}}_k^{tf} s_k^{tf} + e^{tf}_{\text{BS}}. \label{x_cas} 
\end{equation}
where $\bm{\hat{w}}_k^{tf} \in \mathbb{C}^{M_c\times 1}, \forall k\in \mathcal{K}$ is the beamforming vector. $M_c$ and $e^{tf}_{\text{BS}}$ are number of antennas of BS and the quantization noise at BS.

In the sequal, we obtain the received signal and the SINR of DV$_k$ denoted as
\begin{align}
  &\begin{aligned}
    \hat{y}_k^{tf} = (\bm{h}_k^{tf})^H \sum\nolimits_{k\in \mathcal{K}} \bm{\hat{w}}_k^{tf} s_j^{tf} + (\bm{h}_k^{tf})^H e^{tf}_{\text{BS}}+
n_k^{tf},
  \end{aligned}\label{y_cas}\\
  &\hat{\gamma}_k^{tf} = \frac{|(\bm{h}_k^{tf})^H \bm{\hat{w}}_k^{tf}|^2}{\sum\nolimits_{j\neq k} |(\bm{h}_k^{tf})^H \bm{\hat{w}}_j^{tf}|^2 + |(\bm{h}_k^{tf})^H\sigma_{\text{BS}}^2 |^2 +(\sigma_k^{tf})^2}.\label{sinr_cas}
\end{align}

The WTP of CAS in FBL and INFBL regimes is the same as that of CF systems.
Comparing \eqref{x_cas}$\sim$\eqref{sinr_cas} and \eqref{xn}$\sim$\eqref{eq_gamakt}, we find that the distributed antenna cooperation that in CF systems disappears in CAS.  
Unlike the CF systems, CAS is incapable of distributed antenna collaboration due to the centralized antenna deployment. Therefore, it is beneficial to operate user scheduling in different TF resource to achieve diversity gains. As we know, only considering path loss, the theoretical channel capacity difference between CF systems and CAS can be denoted as the following expression related to the user position \cite{Dmimo_CF}.
\begin{equation}
  \Delta (\bm{d})=\text{log}_2\bigg[\sum_{n=1}^N\bigg(\frac{D}{d_n}\bigg)^{\eta_0}\bigg]-\text{log}_2\bigg[\bigg(\frac{D}{d_0}\bigg)^{\eta_0}\bigg]
  \label{delta_d}
\end{equation} 
where $d_n$ stands for the distance between DV and AP$_n$. $\bm{d}=[d_1,\cdots,d_N]$, $d_0$ is the distance between DV and the centralized BS. $D, \eta_0$
represent the reference distance and fading exponent.
\begin{remark}
  It is seen that the system performance of CF systems and CAS is related to the position of users shown in \eqref{delta_d}. Although the performance of the CF system may be worse than CAS (i.e., users are distributed near the center of the system), its capacity coverage is more uniform and easier to achieve better system performance since the distributed antennas more closer to users, 
  especially when users are dispersed throughout the entire system.                

\end{remark}

Simulation results give the numerical comparison of CAS and CF systems in our proposed scenario. 

\section{Numerical results}
In this section, we present numerical results to demonstrate our resource allocation algorithm in CF systems with INFBL regime and FBL regime. And the system performance between CAS and CF systems is analyzed.  
Besides, we compare the resource allocation scheme in MU-MIMO, FU-MIMO and SU-MIMO scenarios. Simulation parameters are given in TABLE \ref{tab:table_para}.
In cell-free MU-MIMO MC systems, APs and UEs are distributed within 500m radius randomly.
The channel coefficients $\{\bm{h}_k^{tf}\},\forall k\in \mathcal{K},\forall t\in \mathcal{T},\forall f\in \mathcal{F}$ are modeled as $\bm{h}_k^{tf}=\eta_k^{tf}\bm{\overline{h}}_k^{tf}$, where $\bm{\overline{h}}_k^{tf}$ is independent and identically distribution (iid) $\sim \mathcal{CN}(0,1)$. The channel gain $\eta _k^{tf}$ equals $\frac{1}{1+(d_{kn}/D)^{\eta_0}}$, where $d_{kn}$ represent the distance between AP$_n$ and UE$_k$,  respectively.    
We denote $N=8, M=2, K=16, \epsilon=10^{-5}, F=3, T=6$, unless otherwise specified. 
\begin{table}[H]
  \renewcommand\arraystretch{1} 
  \caption{Simulation Parameters\label{tab:table_para}}
  \centering
  \setlength{\tabcolsep}{7mm}{
  \begin{tabular}{c|c}
  \hline
  parameter & value\\
  \hline
  cell radius ($r$) & 500m\\
  \hline
  crossover probability ($p_c$)& 0.8 \\
  \hline
  mutation probability ($p_m$) & 0.9 \\
  \hline
  signal to noise (SNR) & 40dB \\
  \hline
  decoding error probability ($\epsilon$) & 10$^{-1}$$\sim$10$^{-9}$\\
  \hline
  number of APs ($N$) & 1,2,4,8,16\\
  \hline
  number of users ($K$)& 1,2,8,16,24\\
  \hline 
  number of subcarriers ($F$)& 1$\sim$5\\
  \hline
  the maximum latency ($T$) & 2,4,6,8,10\\
  \hline
  \end{tabular}
  }
\end{table}

\begin{figure}[htp]
  \centering
  \includegraphics[width=8cm]{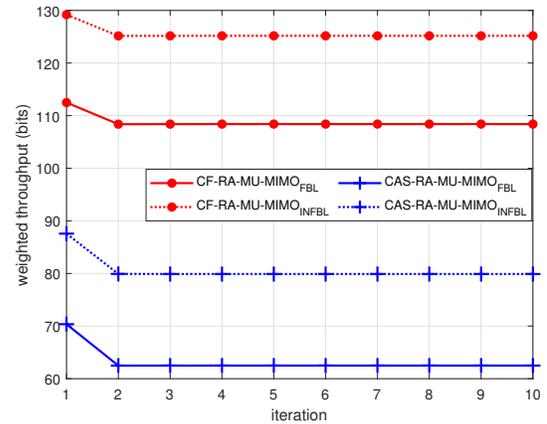}
  \caption{Convergence of the beamforming design in CAS and CF systems with infinite and FBL.}
  \label{fig_conver_alg1}
\end{figure}

Fig. \ref{fig_conver_alg1} shows the convergence of the beamforming (BF) design algorithms, including Algorithm \ref{alg:algGWMMSE} and \ref{alg:algSCA} in INFBL and FBL regimes, respectively. It is seen that the WTP of each scheme converges quickly, almost within 2 iterations. 
Besides, with the same blocklength, the RA-MU-MIMO scheme in CF systems achieves higher performance gain than that in CAS. Similarly, with the same systems, the performance gain of the RA-MU-MIMO scheme in the INFBL regime is better than that in the FBL regime. The detailed performance gains is compared in TABLE \ref{tab:table_gains}.
\begin{table}[H]
  \renewcommand\arraystretch{1} 
  \caption{Comparison of the performance gains\label{tab:table_gains}}
  \centering
  \setlength{\tabcolsep}{7mm}{
  \begin{tabular}{c|c}
  \hline
  {Scenario} & Performance gain\\
  \hline
  CF system & INFBL \textgreater \enspace FBL: 13.41\% \\
  \hline
  CAS sytem&  INFBL \textgreater \enspace FBL: 21.81\%\\
  \hline
  FBL regime & CF \textgreater \enspace CAS: 42.34\%\\
  \hline
  INFBL regime&  CF \textgreater \enspace CAS: 36.14\%\\
  \hline
  \end{tabular}
  }
\end{table}

\begin{figure}[htp]
  \centering
  \includegraphics[width=3.5in]{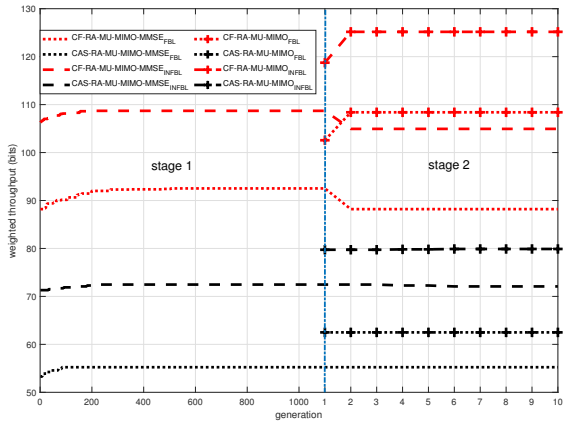}
  \caption{Convergence of the joint user scheduling and beamforming design of Algorithm 1.}
  \label{fig_conver_alg2}
\end{figure}

Fig. \ref{fig_conver_alg2} illustrates the convergence of the joint user scheduling and beamforming design algorithm (Algorithm \ref{alg:algGA}). In each generation of user scheduling, the fitness of each individual is obtained from beamforming design result, where beamforming design algorithm is included in Algorithm \ref{alg:algGA}.
The computational complexity of the proposed beamforming method is much higher than that of the linear MMSE beamformer. Therefore, two-stage algorithm is applied to the Algorithm \ref{alg:algGA}. We first replace the proposed beamforming method with linear MMSE beamformer for calculating the fitness function for each individual in the first stage. Finally, the convergent user scheduling optimization result and the optimal fitness value based on the linear MMSE beamformer are obtained as shown in Fig. \ref{fig_conver_alg2}-stage 1. 
In the second stage, the obtained user scheduling scheme is set as the initial value of user scheduling scheme in the next ten generations of the Algorithm \ref{alg:algGA}, where the proposed beamforming method is adopted in these ten generations. 
As shown in Fig. \ref{fig_conver_alg2}-stage 2, the WPT remains unchanged during the total ten generations both with FBL and INFBL regimes in CAS, while that in CF systems keeps changing within 2 generations, which verifies that the MMSE beamformer instead of the proposed beamforming design method in Algorithm \ref{alg:algGA} can ultimately obtain the same user scheduling optimization result and greatly reduce the computational complexity. Namely, Algorithm \ref{alg:algGA} with proposed beamformer converges within ten generations with a better initial user scheduling scheme obtained from linear and low computational complexity MMSE beamformer pre-used.
\begin{figure}[htp]
  \centering
  \includegraphics[width=3.5in]{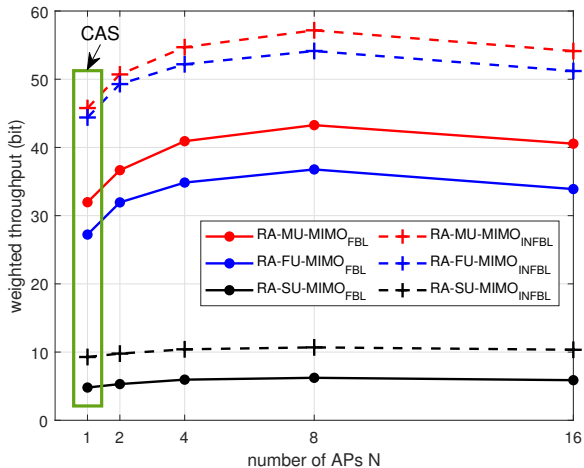}
  \caption{Impact of the number of APs on WTP.}
  \label{fig_AP}
\end{figure} 

In Fig. \ref{fig_AP}, the WPT under different numbers of APs is given to show the impact of the spatial DoF and array gain on WPT. The total number of antennas equals 16. Hence, the number of APs equals 1 (i.e., in CAS) and 16 (i.e., in CF systems) means there are only array gain and spatial DoF, which leads to lower WPT. We find that when the number of APs equals 8, the joint effect of spatial DoF and array gain enables WPT to reach its maximum value. It is seen that the system performance in this scenario is not directly related to the number of APs. It may be related to the trade-off between array gain and spatial DoF. However, the WPT in CF systems is superior to that in CAS. And the RA-MU-MIMO scheme achieves the highest WPT than its comparison schemes both in CAS and CF systems with FBL and INFBL regimes. Similarly, the WPT of each resource allocation scheme in the INFBL regime is higher than its counterpart in the FBL regime due to INFBL and error-free assumption.

\begin{figure}[htp]
	\centering
	\includegraphics[width=3.5in]{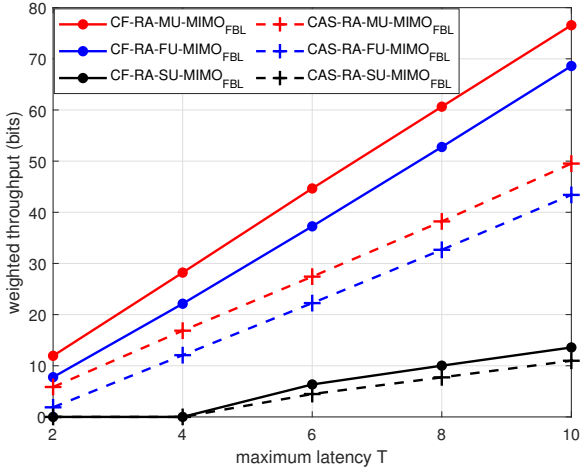}
	\caption{Impact of the maximum latency on WTP in FBL regime.}
	\label{fig_Dlantency}
  \end{figure}

Fig. \ref{fig_Dlantency} studies the impact of the maximum latency on the WPT of three resource allocation methods in CAS and CF systems with FBL. The increasing maximum latency leads to more time-resource being utilized, resulting in higher WPT of all schemes. And the WPT of each scheme in CF systems is superior to its counterpart in CAS, owing to the additional spatial DoF. Under the same maximum latency, RA-MU-MIMO scheme obtains the highest WPT compared to the other two methods, while the RA-SU-MIMO scheme achieves the lowest WPT among the three resource allocation schemes. However, the performance gap between MU-MIMO and FU-MIMO is smaller than that between MU-MIMO/FU-MIMO and SU-MIMO due to the multiuser MIMO diversity gain. Nevertheless, the computational complexity of the RA-MU-MIMO scheme is likely to be reduced compared to the RA-FU-MIMO scheme since the computational complexity of BF-FBL equals $\mathcal{O} ((4K)^3(3KTF+T+K))$ which is related to the number of users. In addition, the WTP of RA-SU-MIMO equals 0 when $T=\{2,4\}$, because the number of users $K=16$ is greater than the total resources elements $T \times F = \{6,12\}$ that is not able to satisfy the QoS of some users in RA-SU-MIMO scheme in which one RE almost serves one user. 

\begin{figure}[htbp]
  \subfigure[CF vs CAS in FBL regime]{
    \centering
      \includegraphics[width=3.5in]{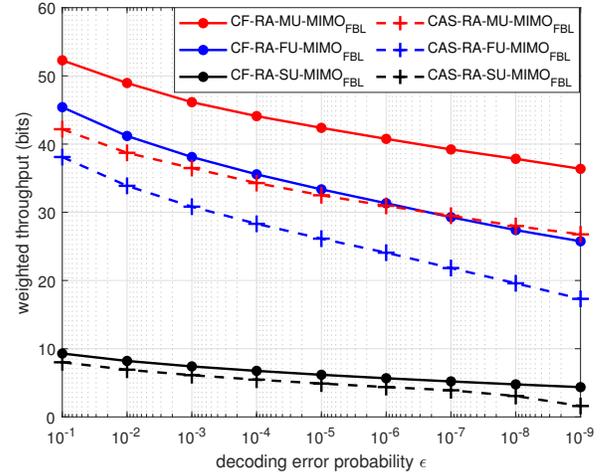}
      \label{CFvsCAS_EPS}
  }
  \subfigure[The trade-off among the WPT, latency and reliability]{
    \centering
      \includegraphics[width=3.5in]{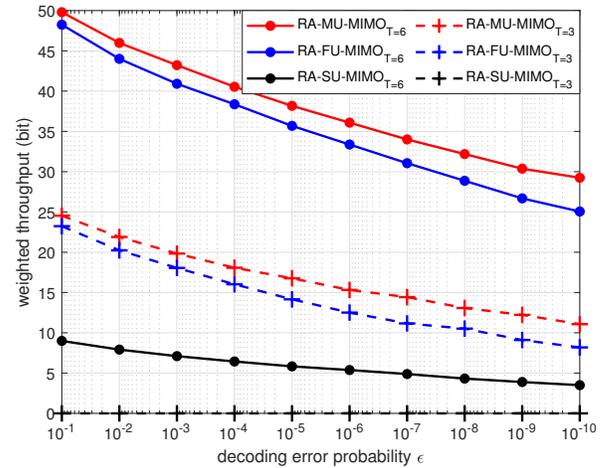}
      \label{epsVsLatency}
  }
  \caption{Impact of the decoding error probability on WTP in FBL regime.}
  \label{fig_Depsilon}
\end{figure}

As shown in Fig. \ref{fig_Depsilon}, we describe the impact of the decoding error probability (DEP) on WPT in FBL regime. For each resource allocation scheme, higher WPT is obtained with a higher DEP, but the reliability of the system is significantly reduced. In Fig. \ref{CFvsCAS_EPS}, we find that the WPT of each scheme in CF system is higher than its corresponding scheme in CAS. 
In Fig. \ref{epsVsLatency}, under the given DEP (i.e., the reliability requirement), higher WTP of each method can be achieved by increasing more number of time slots $T$ (i.e., the maximum latency requirements). 
Therefore, the performance improvement of two of the WPT, reliability, and latency will inevitably lead to the performance degradation of the remaining one, which suggests a performance trade-off among them. 
However, both in Fig. \ref{CFvsCAS_EPS} and Fig. \ref{epsVsLatency}, the performance of the RA-MU-MIMO scheme is superior to that of the counterparts. When the DEP is larger than $10^{-7}$, the WPT of RA-MU-MIMO in CAS starts exceeding that of RA-FU-MIMO in CF systems. Similarly, the RA-SU-MIMO scheme has the lowest WPT under the same DEP compared to the other two schemes. Besides, the superiority of the RA-MU-MIMO scheme appears significantly when the DEP decreases and the time resources are limited. On the one hand, the WPT gap between the RA-MU-MIMO scheme and the RA-FU-MIMO scheme gradully widens with the DEP decreases. On the other hand, the QoS of users in RA-SU-MIMO scheme is unable to be satisfied due to resource deficiency leading to the zero WPT. To  this end, it is necessary to operate RA-MU-MIMO to balance the system performance and limited resources.

\begin{figure*}[htbp]
  \subfigure[CF vs CAS in FBL regime]{
    \centering
      \includegraphics[width=2.33in]{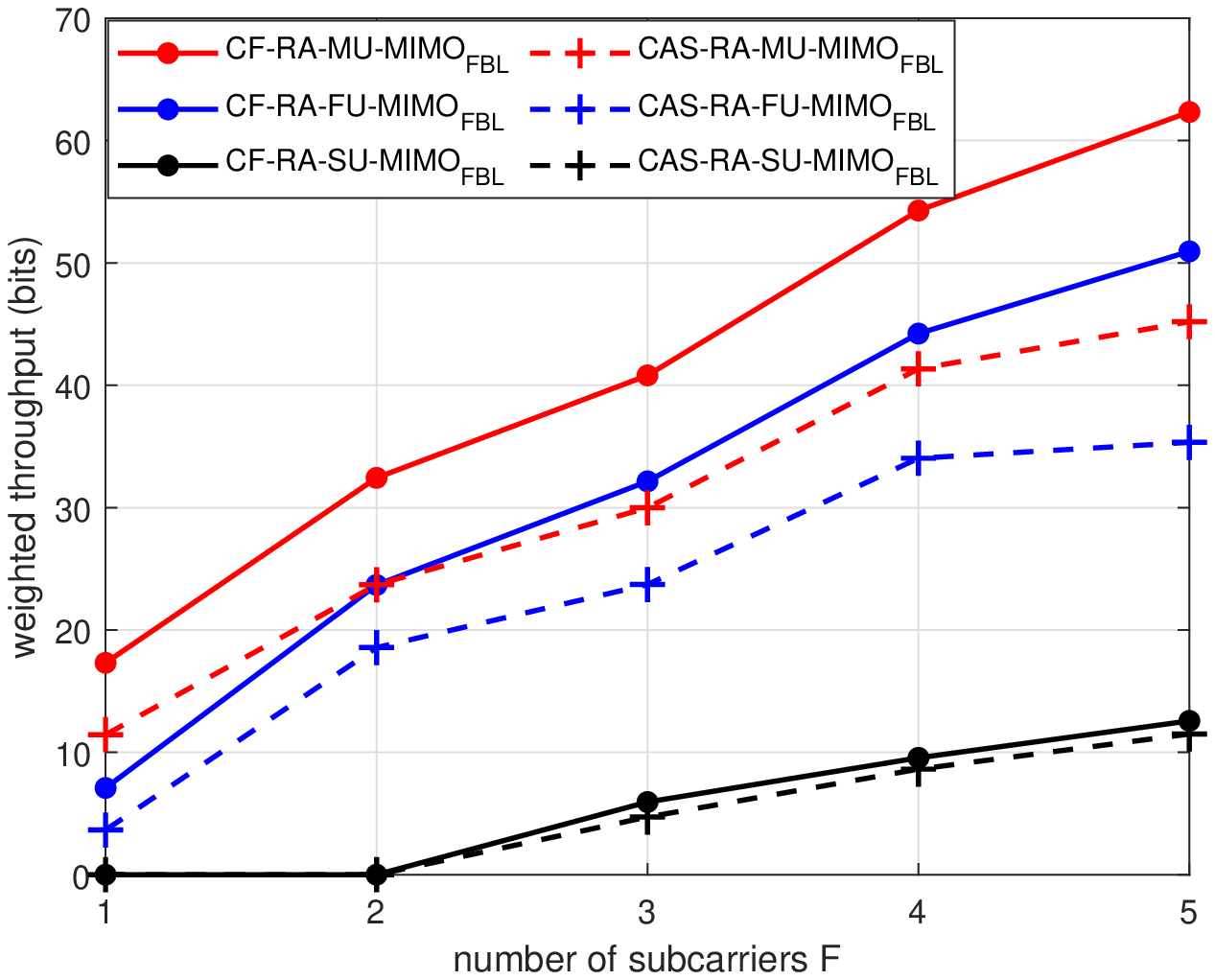}
      \label{CFvsCAS_FBL_F}
  }
  \subfigure[CF vs CAS in INFBL regime]{
    \centering
\includegraphics[width=2.3in]{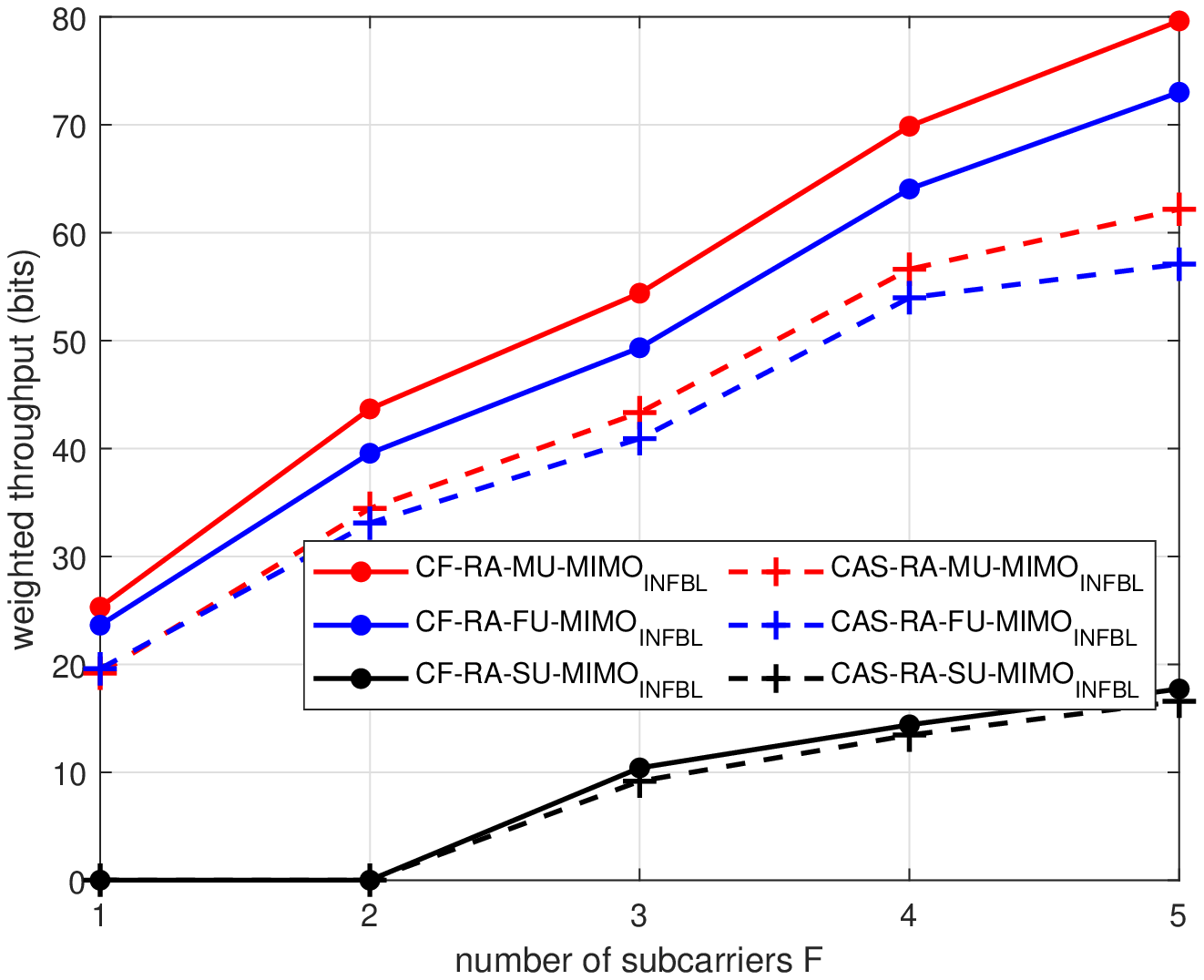}
      \label{CFvsCAS_INFBL_F}
  }
  \subfigure[FBL vs INFBL in CF systems]{
    \centering
\includegraphics[width=2.4in]{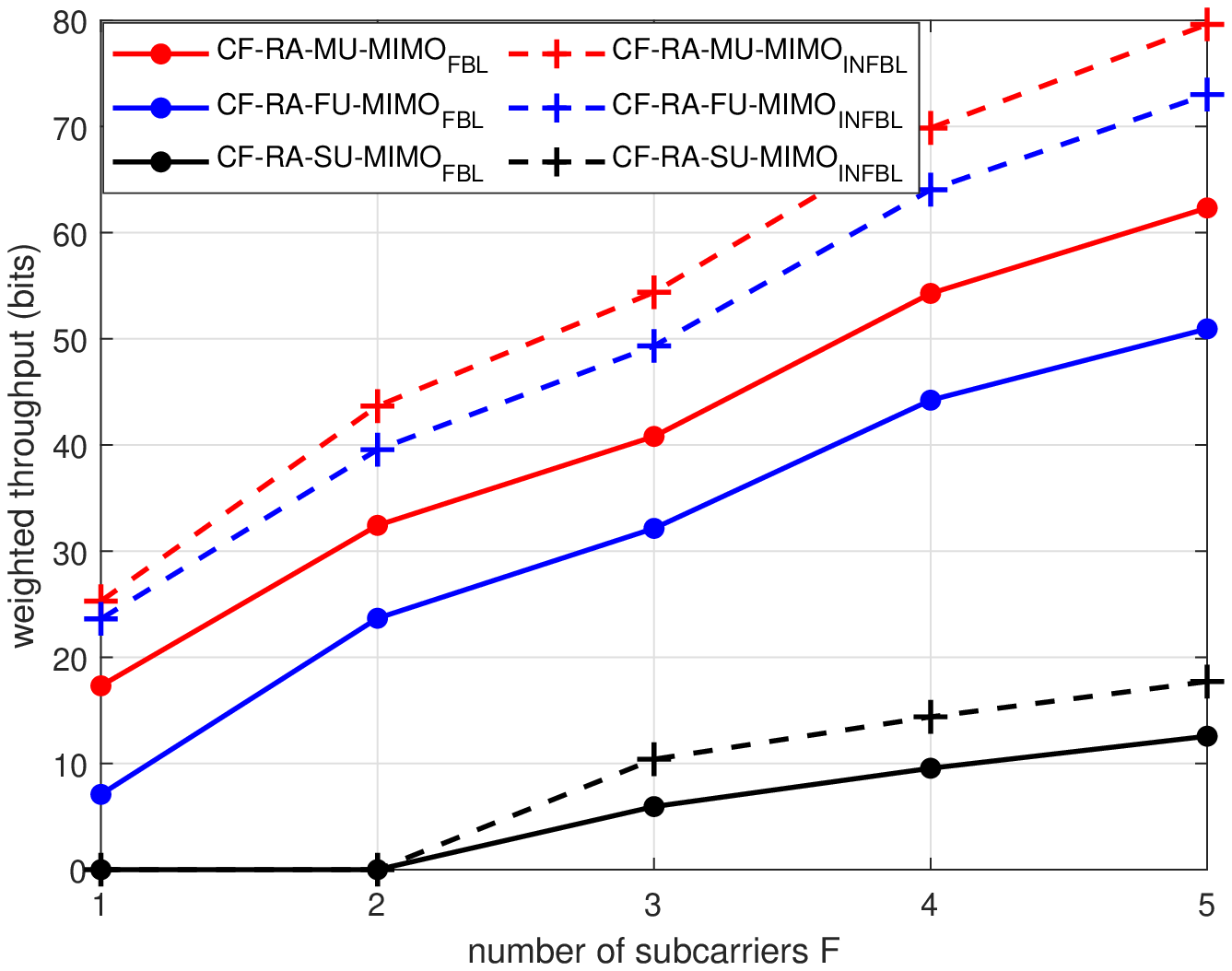}
      \label{FBLvsINFBL_F}
  }
  \caption{Impact of the number of subcarriers on weighted throughput.}
  \label{fig_Dsubcarriers}
\end{figure*}

Fig. \ref{fig_Dsubcarriers} investigates the impact of the number of subcarriers on the WPT. With the increasing of the subcarriers, the WPT of each resource allocation scheme increases as well, which benefits from the frequency diversity. Thereinto, the RA-SU-MIMO scheme acquires the lowest WTP compared to the other two resource allocation schemes both in Fig. \ref{CFvsCAS_FBL_F} $\sim$ Fig. \ref{FBLvsINFBL_F}. What's more, the WPT of RA-SU-MIMO scheme equals 0 when the number of subcarriers equal $F=\{1,2$\}, and the corresponding total REs is $T\times F =\{6,12\}$ which is less than the number of users. Therefore, the QoS of some users is unable to be satisfied resulting in the zero value of the WPT. However, RA-MU-MIMO and RA-FU-MIMO schemes may avoid the situation mentioned in the RA-SU-MIMO scheme due to insufficient resource usage, as multiple or all users are served on each time-frequency resource. Therefore, these two resource allocation schemes have higher resource utilization than the RA-SU-MIMO scheme. Similarly, it is seen that the WPT of each scheme in CF systems is greater than that in CAS due to spatial DoF both in FBL and INFBL regimes shown in Fig. \ref{CFvsCAS_FBL_F} and Fig. \ref{CFvsCAS_INFBL_F}.

\begin{figure*}[htbp]
  \subfigure[CF vs CAS in FBL regime]{
    \centering
      \includegraphics[width=2.3in]{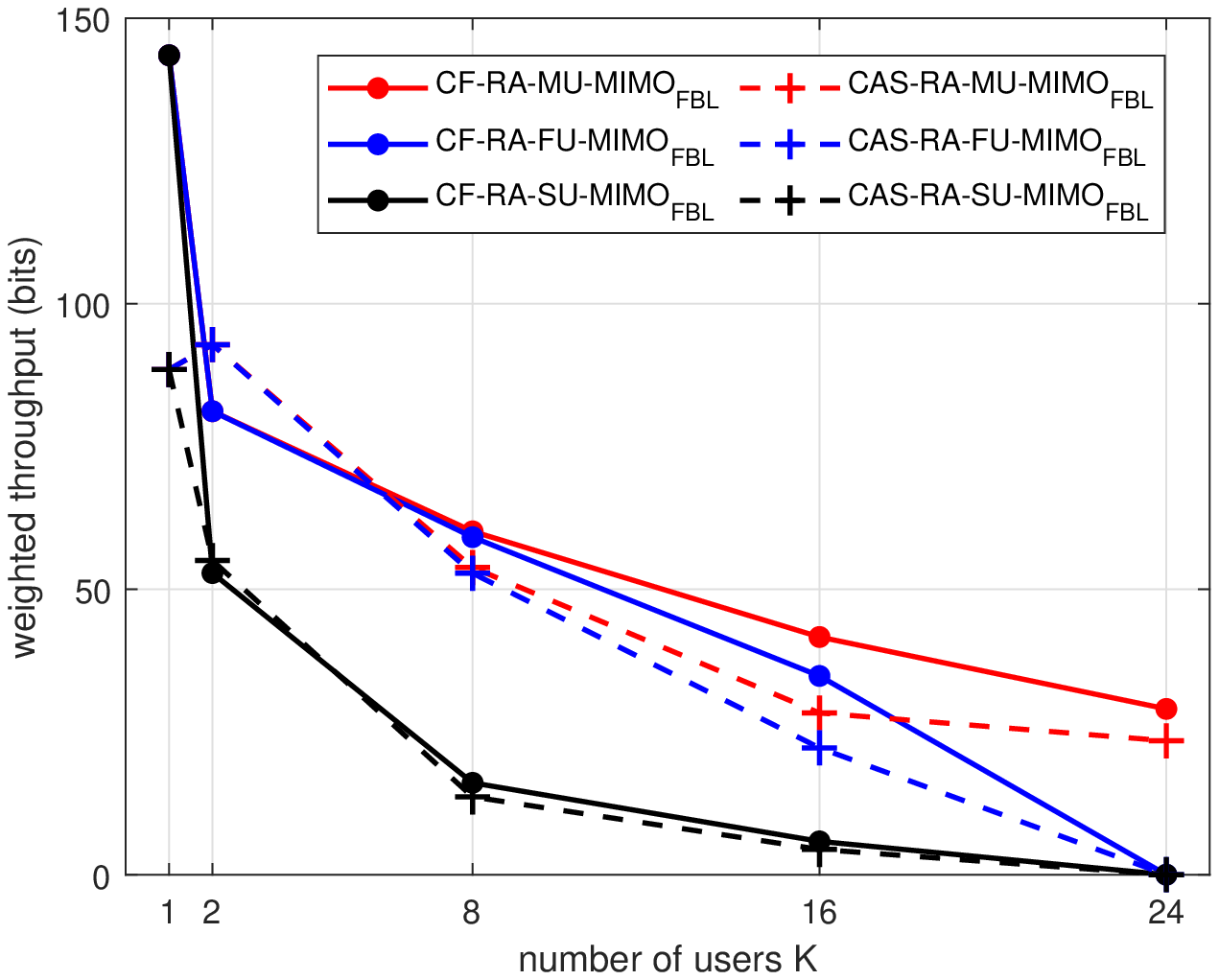}
      \label{CFvsCAS_FBL_UE}
  }
  \subfigure[CF vs CAS in INFBL regime]{
    \centering
\includegraphics[width=2.3in]{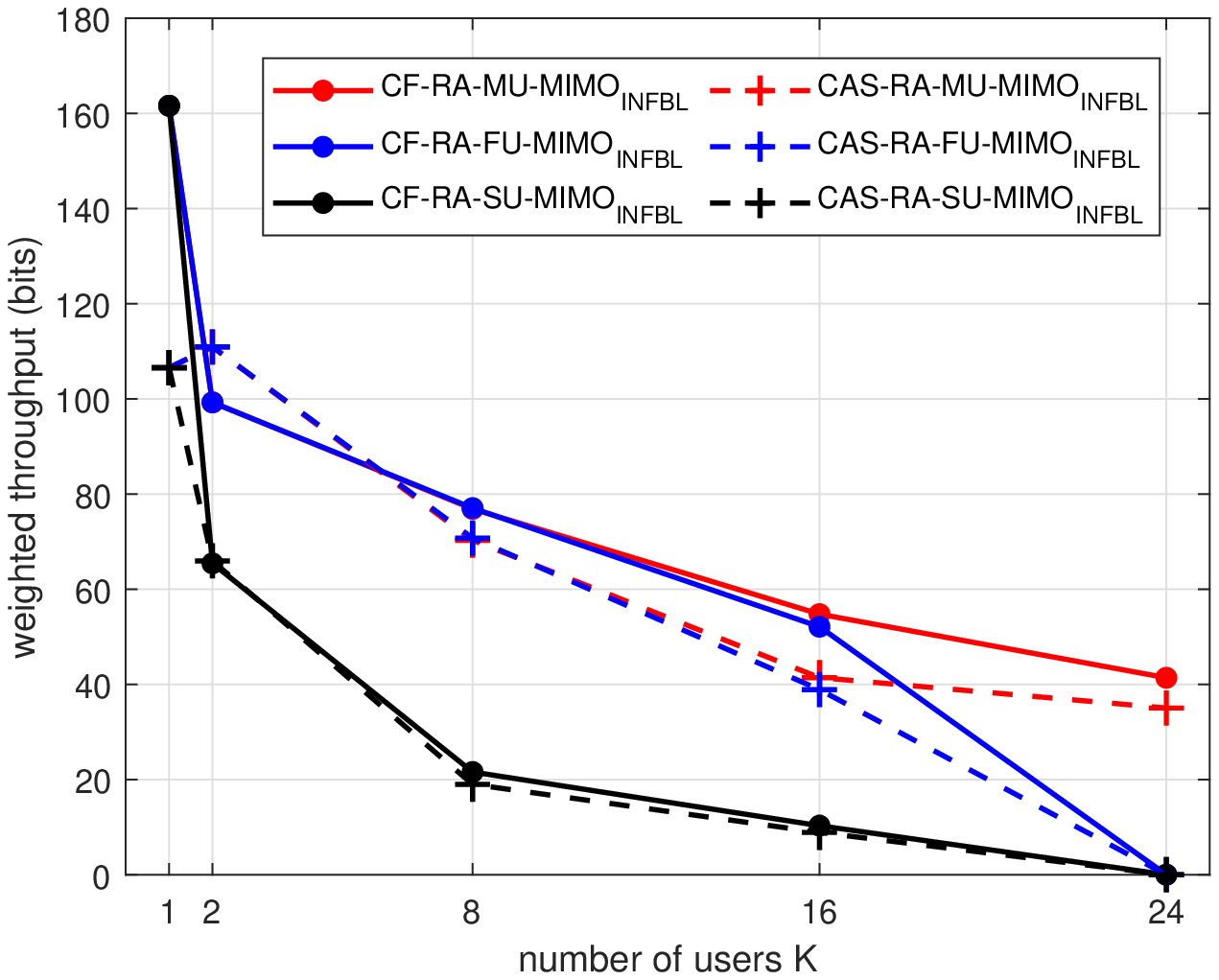}
      \label{CFvsCAS_INFBL_UE}
  }
  \subfigure[FBL vs INFBL in CF systems]{
    \centering
\includegraphics[width=2.3in]{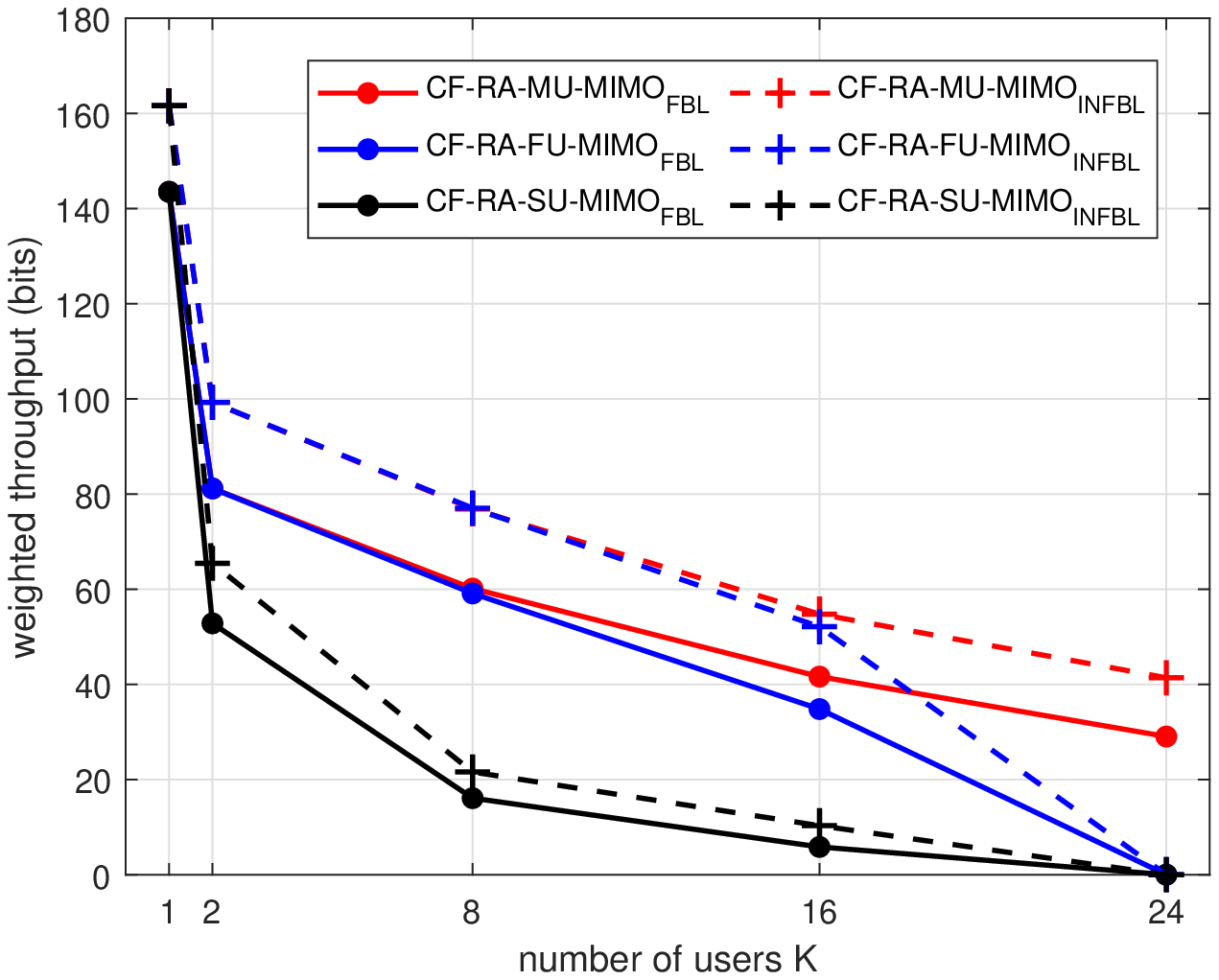}
      \label{FBLvsINFBL_CF_UE}
  }
  \caption{Impact of the number of users on weighted throughput.}
  \label{fig_Dusers}
\end{figure*} 

In Fig. \ref{fig_Dusers}, we further explore the impact of the number of users on WPT. As the number of users increases, the WPT of all resource allocation schemes decreases, beacuse the inter-user interference increases gradually with increasing users. Furthermore, the resources are allocated to more users with diverse channel gain which leads to system performance degradation. 
Besides, when the number of users equals 1 (i.e., $K$=1), three resource allocation schemes in CF systems achieve almost the same performance gain. When the number of antennas is larger than the number of users (i.e., $K$\textless
16), the RA-MU-MIMO and RA-FU-MIMO achemes obtain the same WPT. Meanwhile, the WPT of RA-MU-MIMO and RA-FU-MIMO schemes in CAS exceeds that in CF systems shown in Fig. \ref{CFvsCAS_FBL_UE} and Fig. \ref{CFvsCAS_INFBL_UE} since the array gain may be greater than the spatial DoF in this situation.
As the number of users approaches to the number of antennas, the WPT of RA-MU-MIMO scheme is better than that of RA-FU-MIMO scheme both in CAS and CF systems with FBL and INFBL. And the WPT of each resource allocation scheme in CF systems is higher than its counterpart in CAS. To furthe increase the number of users (i.e., the number of users is greater than that of the antennas and REs), it is seen  that not only the RA-SU-MIMO scheme but also the RA-FU-MIMO scheme are unable to satisfy the QoS of each user while the system performance of the RA-MU-MIMO scheme may still steadily decrease with the increased number of users. 
\vspace{0.5cm}
\begin{figure*}[htbp]
  \subfigure[CF vs CAS in FBL regime]{
    \centering
      \includegraphics[width=2.3in]{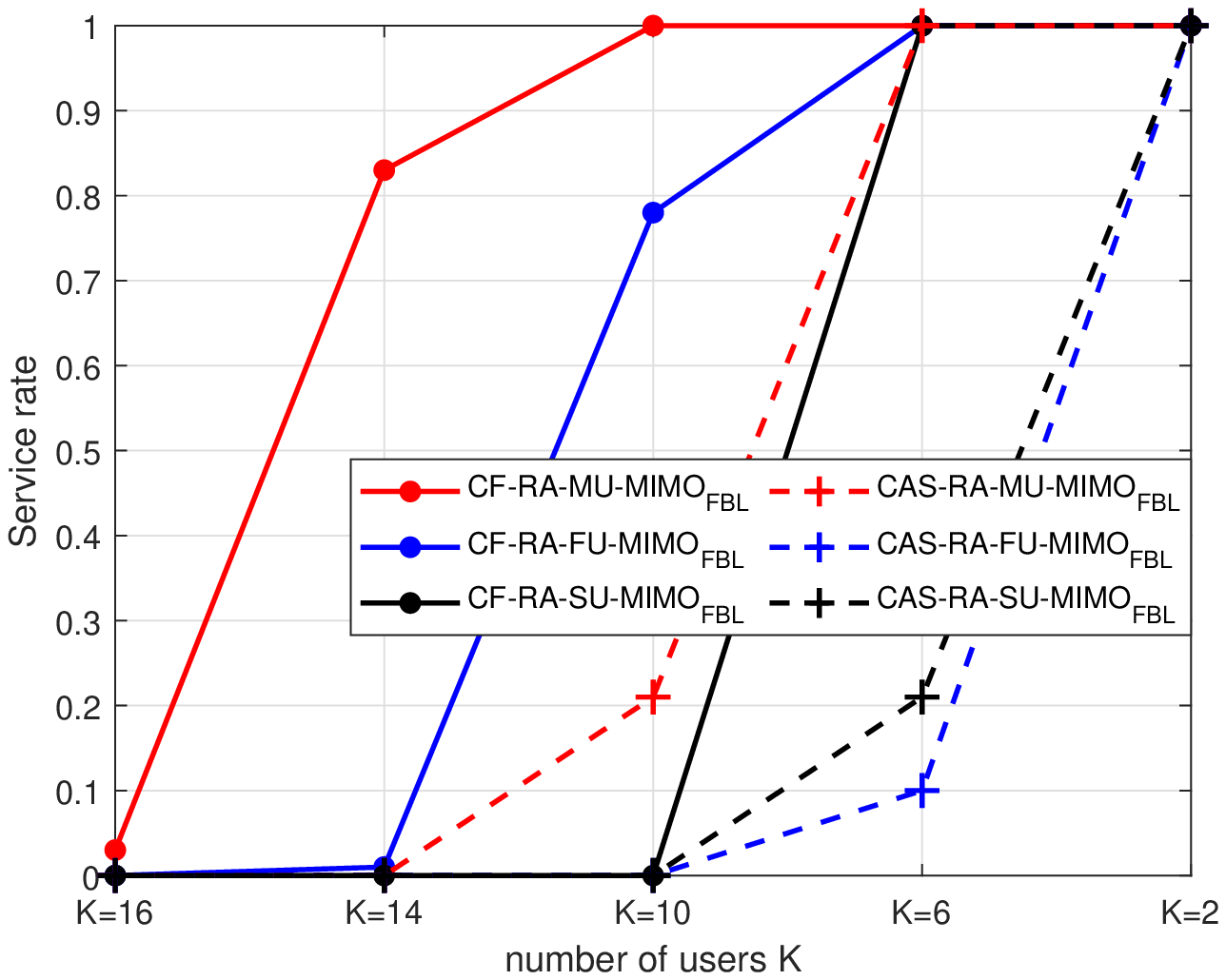}
      \label{CFvsCAS_FBL_sr}
  }
  \subfigure[CF vs CAS in INFBL regime]{
    \centering
\includegraphics[width=2.3in]{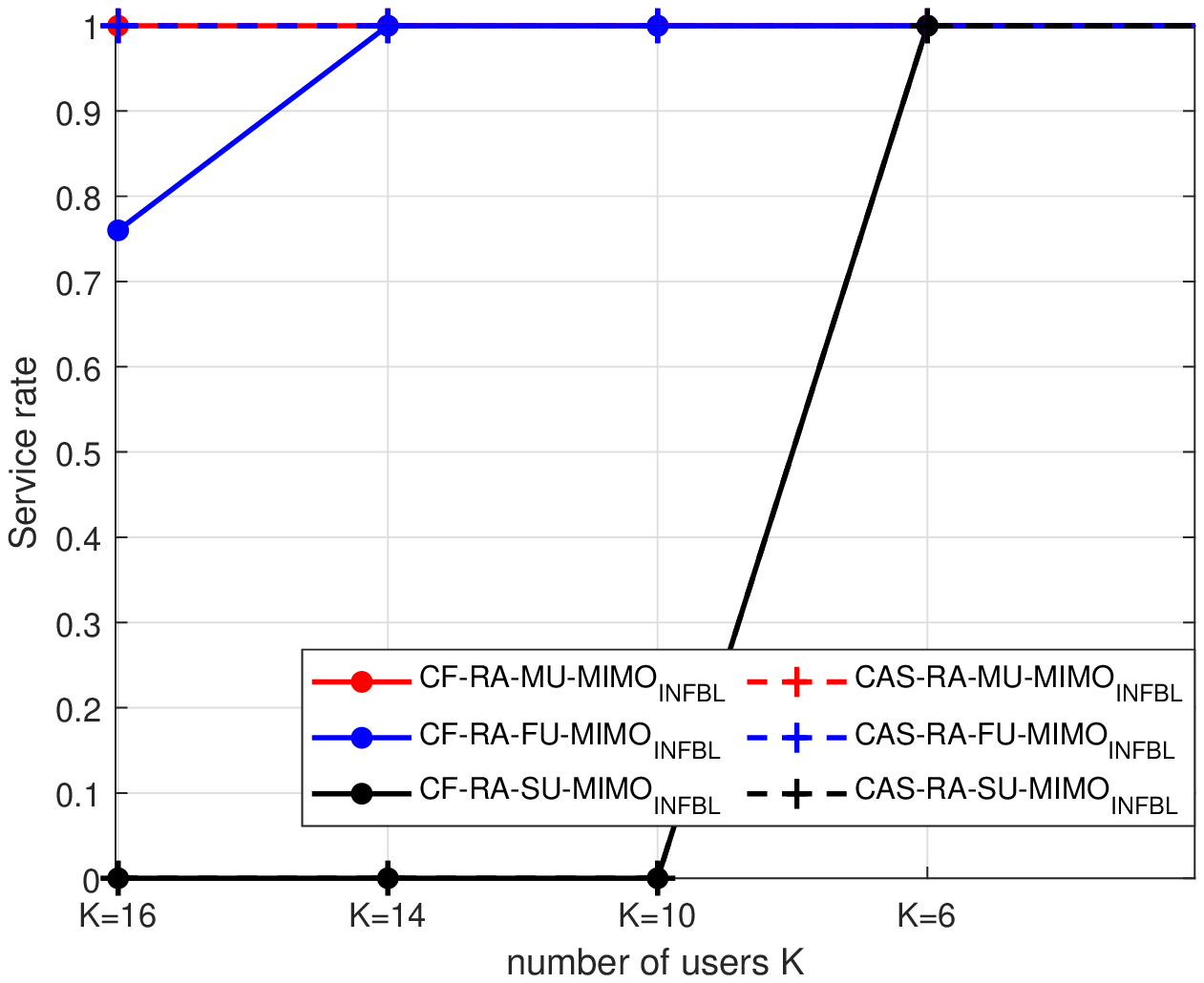}
      \label{CFvsCAS_INFBL_sr}
  }
  \subfigure[FBL vs INFBL in CF systems]{
    \centering
\includegraphics[width=2.3in]{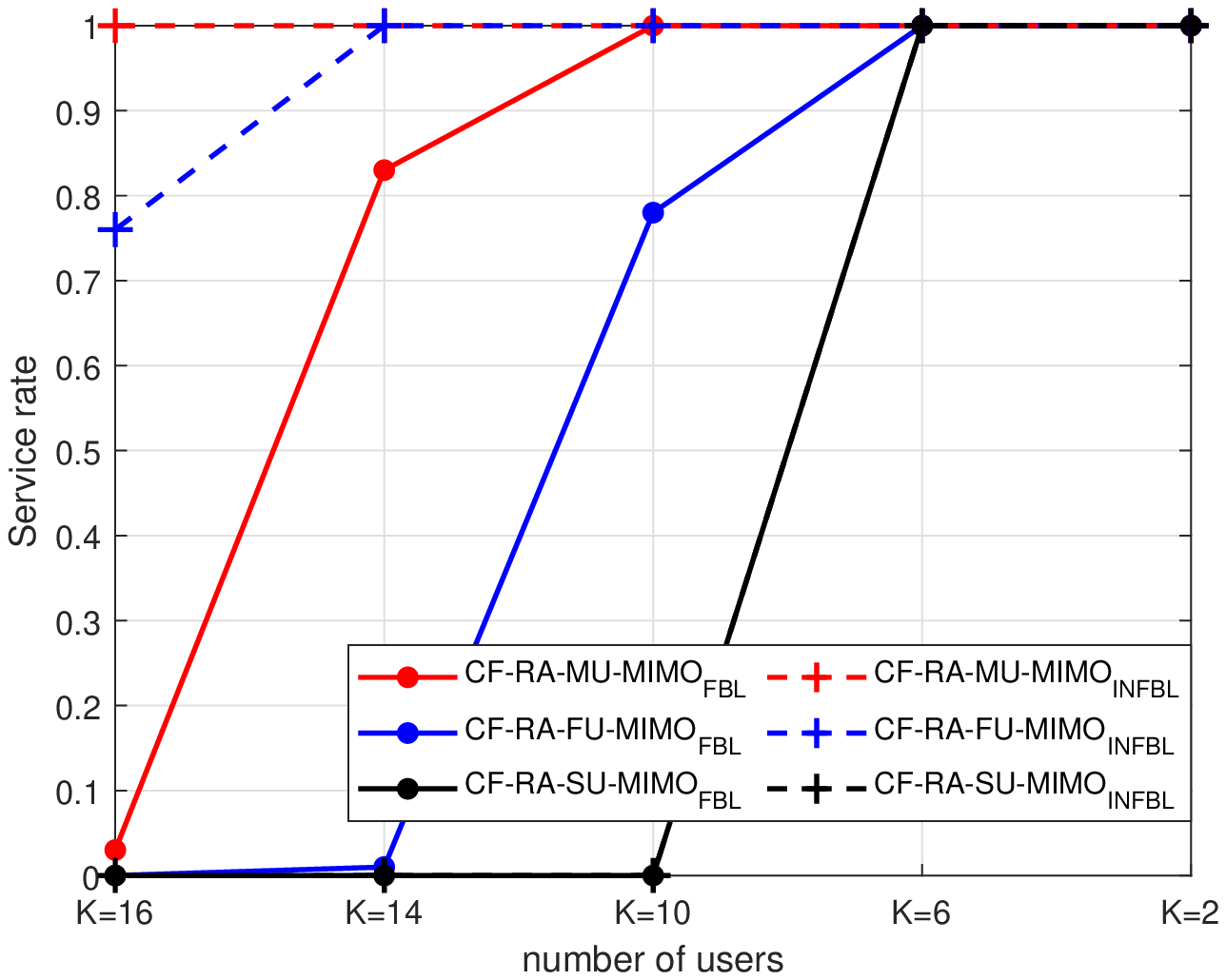}
      \label{FBLvsINFBLCF_sr}
  }
  \caption{Service rate of each resource allocation in CAS and CF systems with FBL and INFBL.}
  \label{fig_Dservice}
\end{figure*}  
In Fig. \ref{fig_Dservice}, we discuss the service rate (SR) of each resource allocation scheme under different numbers of users both in CAS and CF systems with FBL and INFBL, where SR represents the proportion of the number of Monte Carlo simulations that meet the QoS for all users to the overall number of Monte Carlo simulations. 
The total resource elements equals $T\times F = 6$, the number of APs in CF systems is 6 and the number of antennas per AP is 2 (i.e., the total number of antennas is 12). As shown in the figure, when the number of users exceeds the number of resource elements, the SR of users in RA-SU-MIMO equals zero since the resource elements are not adequate for each user without inter-user interference. And the SR in the RA-FU-MIMO scheme in CAS with FBL is even lower than that of RA-SU-MIMO which may be caused by the severe inter-user interference. Furthermore, when the number of antennas is less than the number of users, the SR of users approaches zero for all of the resource allocation schemes since the spatial DoF is lacking for the increasing number of users. However, the SR of the RA-MU-MIMO scheme is greater than that of the counterparts under the same number of users, which means the importance of resource allocation. Besides, the SR of users in the INFBL regime is superior to that in its corresponding FBL regime in Fig. \ref{FBLvsINFBLCF_sr} due to INFBL and error-free assumption. In addition, the SR of each resource allocation scheme in CF systems with INFBL is the same as that in CAS, while the SR of each resource allocation scheme in CF systems with FBL is better than that in CAS, which indicates that the system performance in FBL regime may be more sensitive than that in INFBL regime due to the impact of channel dispersion.

\section{Conclusion\label{sec: conclusion}}
In this paper, we investigated the maximization problem of WTP in a CF MU-MIMO MC system both with FBL and INFBL, where the constraints of the total system power consumption and the minimum QoS of each user are satisfied. 
We jointly optimized the user scheduling scheme and the beamformer to maximize the system performance in INFBL regime and FBL regime under the given BLER and the maximum latency. 
Since the problem is a MINLP problem which is highly non-convex, a nested iteration algorithm (USBDA) based on GA and SCA was proposed to efficiently solve this intractable issue. To step further, a two-stage algorithm was used for the USBDA algorithm to reduce the computational complexity which was verified in the Section V. Besides, simulation results
verified that CF architecture provides more spatial DoF to achieve higher spectral efficiency compared to the CAS. In addition, the proposed RA-MU-MIMO resource allocation scheme is superior to the other two comparison resource allocation schemes, including RA-SU-MIMO and RA-FU-MIMO resource allocation schemes, resulting in higher WTP with limited resources certainly both in CAS and CF systems. 
What's more, we found that the proposed resource allocation scheme in the FBL regime is more suitable for URLLC scenarios where the latency and reliability are strictly met, while that in the INFBL regime is more satisfied for eMBB scenarios with infinite blocklength and error-free assumptions. 
Furthermore, more TF resources and antennas lead to higher system performance due to multiplexing gain and diversity gain. And the system performance in CF systems is superior to that of its collocated counterpart as the number of users continuously increases due to the more spatial DoF in CF systems. Nevertheless, when the number of users is quite small, the system performance in CAS is better than CF systems resulting from the array gain.




\end{document}